\documentclass[preprint,12pt]{elsarticle}
\usepackage{xcolor}
\pagecolor{white}
\usepackage{hyperref}
\usepackage{subcaption}

\usepackage{wrapfig}
\usepackage[section]{placeins}
\graphicspath{{figures/}}
\usepackage{amssymb}
\usepackage{amsthm, physics, bm}
\usepackage{amsmath}

\usepackage{lineno}

\journal{Nuclear Inst. and Methods in Physics Research, A, }

\begin{document}
\begin{frontmatter}


\title{Accurate Determination of the Electron Spin Polarization In Magnetized Iron and Nickel Foils for M\o ller  Polarimetry }


\author[a]{D. C. Jones\corref{cor1}}
\ead{donald.jones@temple.edu, jonesdc@jlab.org}
\author[a]{J. Napolitano}
\address[a]{Temple University, Philadelphia, PA, 19122}
\author[b]{P. A. Souder}
\author[a,c]{D. E. King}
\address[b]{Syracuse University, Syracuse, NY 13244}
\author[c]{W. Henry}
\author[c]{D. Gaskell}
\address[c]{Jefferson Lab, Newport News, VA 23606}
\author[d]{K. Paschke}
\address[d]{University of Virginia, Charlottesville, VA 22903}
\cortext[cor1]{corresponding author}

\begin{abstract}
The M\o ller polarimeter in Hall A at Jefferson Lab in Newport News, VA, has provided reliable measurements of electron beam polarization for the past two decades. Past experiments have typically required polarimetry at the 1\% level of absolute uncertainty which the M\o ller polarimeter has delivered. However, the upcoming proposed experimental program including MOLLER and SoLID have stringent requirements on beam polarimetry precision at the level of 0.4\%\cite{MOLLER2014, SoLID2019}, requiring a systematic re-examination of all the contributing uncertainties. 

M\o ller polarimetry uses the double polarized scattering asymmetry of a polarized electron beam on a target with polarized atomic electrons. The target is a ferromagnetic material magnetized to align the spins in a given direction. In Hall A, the target is a pure iron foil aligned perpendicular to the beam and magnetized out of plane parallel or antiparallel to the beam direction. The acceptance of the detector is engineered to collect scattered electrons close to 90$^{\circ}$ in the center of mass frame where the analyzing power is a maximum (-7/9).  

One of the leading systematic errors comes from determination of the target foil polarization. Polarization of a magnetically saturated target foil requires knowledge of both the saturation magnetization and $g^\prime$, the electron $g$-factor which includes components from both spin and  orbital angular momentum from which the spin fraction of magnetization is determined. Target foil polarization has been previously addressed in a 1997 publication ``A precise target for M\o ller polarimetry" by deBever {\it et. al} \cite{deBever1997} at a level of precision sufficient for experiments up to this point. Several shortcomings with the previous published value require revisiting the result prior to MOLLER. This paper utilizes the existing world data to provide a best estimate for target polarization for both nickel and iron foils including uncertainties in magnetization, high-field and temperature dependence, and fractional contribution to magnetization from orbital effects. We determine the foil electron spin polarization at 294~K to be 0.08020$\pm$0.00018 (@4~T applied field) for iron and 0.018845$\pm0.000053$ (@2~T applied field) for nickel. We conclude with a brief discussion of additional systematic uncertainties to M\o ller polarimetry using this technique.
\end{abstract}


\begin{highlights}
\item Magnetization of Fe and Ni at room temperature from world data
\item Spin fraction of magnetic moment for Fe and Ni from world data
\end{highlights}

\begin{keyword}



\end{keyword}

\end{frontmatter}


\section{Introduction to M\o ller polarimetry}
M\o ller polarimetery utilizes the analyzing power of polarized electron-electron scattering to determine the polarization of an electron beam. The polarized target is usually composed of iron or a highly ferromagnetic material. Elastically scattered events (beam electrons from atomic electrons) produce back-to-back electrons in the center of mass frame. If both are detected in coincidence background contributions can be significantly reduced.

Following the analysis in \cite{Swartz1995}, where the center of mass energy of the $e^-e^-$ pair $E_{CM}\gg m_e$, M\o ller scattering at tree level in the electron-electron center of mass (CM) system is given by
\begin{align}
\frac{d\sigma}{d\Omega_{cm}}=&\frac{\alpha^2}{E_{CM}^2}\frac{\left(3+\cos^2\theta\right)^2}{\sin^4\theta}\bigg[1- \bigg. \nonumber \\
&\bigg. P^{\rm targ}_{\ell}P^{\rm beam}_{\ell}A_{\ell}(\theta)-P^{\rm targ}_tP^{\rm beam}_tA_t(\theta)\cos\left(2\phi-\phi_{\rm beam}-\phi_{\rm targ}\right)\bigg]
\label{eq:moller_cx}
\end{align}
where the subscripts $t$ and $\ell$ refer to transverse and longitudinal polarization respectively. The CM scattering angle is $\theta$ and $\phi$ is the azimuthal angle of the scattering plane. $phi_{\rm beam(targ)} $ is the azimuthal angle of the transverse beam(target) polarization. The analyzing powers for longitudinal and transverse polarization are given by
\begin{equation}
A_{\ell}(\theta)=\frac{\left(7+\cos^2\theta\right)\sin^2\theta}{\left(3+\cos^2\theta\right)^2}~~~\textrm{and}~~~A_t(\theta)=\frac{\sin^4\theta}{\left(3+\cos^2\theta\right)^2}.
\label{eq:analyzing_pow}
\end{equation}
At $\theta=90^\circ$, $A_{\ell}$ is at its maximum value of 7/9 which is a factor of 7 larger than $A_t$ giving M\o ller polarimetery much more sensitivity to longitudinal polarization. The optics of the M\o ller polarimeter in Hall A are tuned to accept events near this maximum analyzing power for longitudinal polarization. The M\o ller polarimeter in Hall A with its Fe foil polarized ``out of plane" in the beam direction ($P^{\rm targ}_t=0$) is designed to measure the longitudinal polarization and be insensitive to the transverse polarization. Nevertheless, if the foil or magnetizing coils are not properly aligned and a transverse foil polarization develops, a non-negligible component of transverse asymmetry could in principle arise. In the ensuing discussion it will be assumed that the foil is properly aligned such that $P^{\rm targ}_t=0$ and this term will be neglected.\footnote{We can approximate the relative size of this term to justify our neglect of it. Longitudinal polarization at JLab can be adjusted for experiments to within $\pm2^\circ$ of uncertainty, leaving a maximum $P^{\rm targ}_t$ of 0.035. Assuming an anomalously large transverse component of the target polarization due to misalignment of 5\% and a transverse analyzing power that is approximately 1/7 that of the longitudinal gives a maximum transverse polarization contribution (i.e. for a beam and target polarization at the same azimuthal angle) that is 0.025\% that of the longitudinal term.}

Integrating the cross section over the acceptance of the detector gives 
\[
\sigma \propto1-P^{\rm targ}_{\ell}P^{\rm beam}_{\ell}A_{zz},
\]
where $A_{zz}=\langle A_l(\theta)\rangle$, the acceptance-weighted analyzing power. We can now see that the left-right scattering asymmetry $A_{LR}$ is then given by 
\begin{equation}
A_{LR}=\frac{\sigma_R-\sigma_L}{\sigma_R+\sigma_L}=P^{\rm targ}_{\ell}P^{\rm beam}_{\ell}A_{zz},
\label{eq:A_LR}
\end{equation}
where $\sigma_{L(R)}$ are the cross sections for left (right) helicity electrons. Implicit in this form is the assumption that $P_\ell^{\rm beam}$ is the same for both helicity states. If $A_{zz}$ and the target polarization $P_{\ell}^{\rm targ}$ are known, the beam polarization can be determined from the measured scattering asymmetry. 

In the approximation where the target electrons are at rest and the beam energy is large compared to the electron rest mass $m_e$, the relationship between the lab momentum of the scattered electron, $p^{\prime}$, and the center of mass scattering angle $\theta$ is given by 
\begin{equation}
p^{\prime}=\frac{p_b}{2}\left(1+\cos\theta \right),
\label{eq:pvstheta}
\end{equation}
where $p_b$ is the electron beam momentum. Thus momentum analyzing the M\o ller scattered electrons also analyzes in $\theta$. Single arm M\o ller polarimeters leverage this characteristic to reduce potentially overwhelming backgrounds arising from Mott scattering from the nucleus. Using a narrow aperture in $\phi$ to select the scattering plane and a dipole to momentum analyze the scattering events perpendicular to the scattering plane produces a characteristic M\o ller ``stripe" downstream of the dipole. Converting to the lab scattering angle and in the absence of other focussing optics, and using the small angle approximation yield the following relationship between $\theta_{\rm Lab}$ and  momentum:
\begin{equation}
\theta^2_{\rm Lab}=2m_ec\left(\frac{p_b-p^{\prime}}{p^{\prime}p_b}\right).
\label{eq:theta_lab}
\end{equation}
\subsection{The M\o ller polarimeter in Hall A at Jefferson Lab}
Part of the standard equipment in Hall A at Jefferson Lab is the M\o ller polarimeter, used to measure the electron beam polarization in the Hall. Most experiments in the past have had polarization requirements at the several percent uncertainty level easily attained by the M\o ller. Two recent experiments, PREX-2\cite{PREX2021} and CREX, have reached $<$0.9\% uncertainty for M\o ller polarimetry. However,  MOLLER and SoLID, the future parity violation experiments planned for Hall A in 2025 and beyond, require uncertainty in electron polarization at $\pm$0.4\%, a record-breaking level of precision that requires re-examination of all the possible sources of systematic error. This paper is designed to address specifically the uncertainty associated with target foil polarization for these experiments, but has obvious value for other M\o ller polarimeters around the world. Where appropriate, we will provide the means to extrapolate these results to other polarimeters with different designs and operating parameters.

\begin{figure}[ht]
\centering
\includegraphics[width=0.9\textwidth]{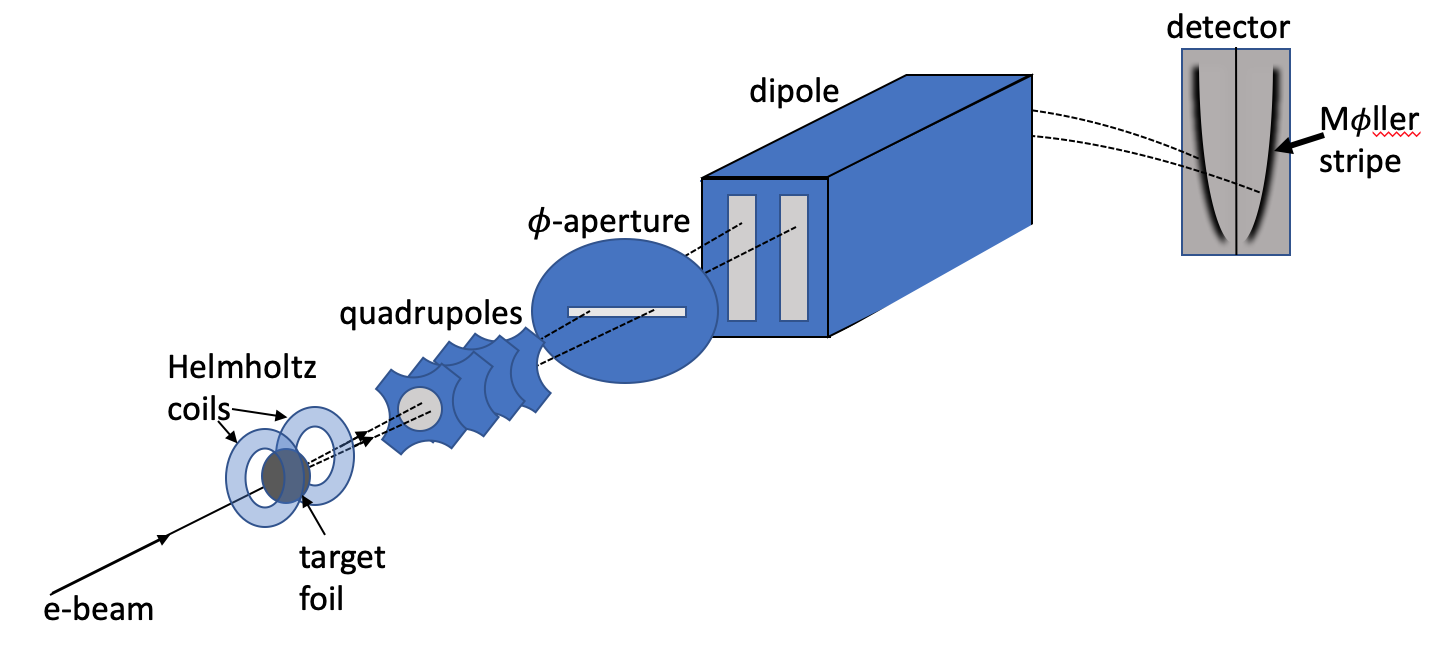}
\caption{\label{fig:moller_diag}Simplified schematic showing the key features of the M\o ller polarimeter setup in Hall A. The electron beam scatters from a polarized foil target. Quadrupole magnets then focus the events of interest through the dipole magnet. An aperture at the front of the dipole limits the $\phi$-acceptance, defining a horizontal scattering plane. Two left-right symmetric narrow vertical apertures in the dipole set the $\theta$ acceptance. The dipole momentum analyzes the scattered electron pairs bending them down onto the detector plane producing characteristic M\o ller stripes.}
\end{figure}

The polarimeter in Hall A is designed to take advantage of both the dipole momentum selection and the coincidence of dual arm detection to further reduce backgrounds. A simple schematic of the Hall A polarimeter is shown in Fig. \ref{fig:moller_diag} illustrating the key features. This polarimeter design adds to the essential elements 4 quadrupoles and an additional horizontal constraint due to the narrow apertures through the dipole. The quadrupoles are used to focus a distribution of M\o ller pairs roughly symmetric about the 90 degree center of mass through the dipole onto the detector. The additional focusing of the quadrupoles inverts the expected typical quadrature curvature (see Eq. \ref{eq:theta_lab}) of the M\o ller stripe on the detector plane as illustrated in Fig. \ref{fig:moller_diag}.

\section{Foil Target Polarization}
In the context of M\o ller polarimetry, the target polarization is produced using a strong magnetic field to align electron spins in ferromagnetic materials. The M\o ller polarimeter target in Hall A consists of a set of thin foils mounted on a target ladder and magnetized out of plane parallel (or anti-parallel) to the beam trajectory by a set of superconducting Helmholtz coils. The superconducting magnet used to polarize the target foils was built by American Magnetics Inc. The field at the center of the coils is horizontal and along the beam-line axis. The maximum field at the center is rated at 5~T, although we do not typically run above 4~T. 

The three ferromagnetic elements, Fe, Co and Ni are the obvious choices for foil targets due to their relatively high magnetization and the precision with which their magnetic properties are known. A list of the main properties of these elements is given in Table \ref{tab:ferro_elem}.
\begin{table}[]
\caption{\label{tab:ferro_elem}Properties of the three ferromagnetic elements. This manuscript focusses on the absolute uncertainties on $M_0$ and $g^\prime$.}
\begin{center}
\begin{tabular}{|r|l|l|l|}\hline
~&Fe&Co&Ni\\\hline
Z&26&27&28\\
Atomic Mass ($\mu$)&55.845(2)&58.933194(4)&58.6934(4)\\
Electron Configuration&[Ar]$4s^23d^6$&[Ar]$4s^23d^7$&[Ar]$4s^23d^8$ \\
Unpaired Electrons&2.2&1.72&0.6\\
Density near r.t. (g/cm$^3$)&7.874&8.900&8.902\\
$M_0$ at 0 K (emu/g)&222&164&58.6\\
$g^{\prime}$&1.92&1.85&1.84\\
Curie Temperature (K)& 1043&1400&631\\\hline
Stable Isotopes & $^{54}$Fe (5.85\%)& $^{59}$Co (100\%)&$^{58}$Ni (68.08\%)\\
~&$^{56}$Fe (91.75\%)&~&$^{60}$Ni (26.22\%)\\
~&$^{57}$Fe (2.12\%)&~&$^{61}$Ni (1.14\%)\\
~&$^{58}$Fe (0.28\%)&~&$^{62}$Ni (3.64\%)\\
~&~&~&$^{64}$Ni (0.93\%)\\\hline
\end{tabular}
\end{center}
\end{table}
The saturation magnetization of Fe and Ni are both known to high accuracy ($\sim0.2\%$), but the low Curie temperature of Ni makes it susceptible to large (percent level) corrections from target heating effects. There are fewer published measurements of high precision on Co than on the other two ferromagnetic elements.

M\o ller polarimetry requires finding the average target electron polarization which is most accurately known at magnetic saturation when further polarization is negligible with increases in applied field. Determining the target polarization requires knowing the magnetization of the target material. Magnetization, $\bf M$, is defined as the magnetic dipole moment per unit volume or in certain contexts, per unit mass. The magnetization provides the magnetic field contributed by a material and relates the flux density $\bf B$ to the auxiliary field $\bf H$ as follows:
\[
\bf{B}=\bf{H}+4\pi\bf{M}.
\]
Note that this is in Gaussian units which are used throughout this document.

While knowledge of magnetization is key to determining target polarization, it includes contributions of both the orbital and spin magnetic moments. Since we only want the spin component we need to find the fraction of the magnetization that comes from spin. This is typically determined from precise measurements of the gyromagnetic ratio (the ratio of a material's magnetization to its angular momentum) of an elemental sample. Thus, the final error on the target polarization will include uncertainties on both the determination of magnetization and of the spin fraction.

In the following sections we look at each of the three elements and determine the systematic uncertainty associated with using each as a target materials. The primary issues to be dealt with are follows:
\begin{itemize}
\item{From 1930-1980 many precise measurements have been made of the magnetization and gyromechanical properties of these elements; however, they do not necessarily agree within error. Sometimes the errors quoted are not realistic given the systematic disagreement in the data. The sources of systematic difference are often not known and yet results are averaged together and the final error estimated from the variance of the data.}
\item{No mention is made of the nuclear contribution to the magnetic moment. The nuclear magneton is smaller than the Bohr magneton by a factor of $m_e/m_p\sim0.05\%$. Fortunately, the main isotopes that make up iron and nickel are even-even and have spinless nuclei, but for Co the average is 4.6 nuclear magnetons making the contribution potentially above the 0.1\%.}
\item{Measurements of magnetization and gyromechanical properties are not made at the same applied field and temperature where the M\o ller polarimeter operates, necessitating corrections to account for these differences. The corrections must be known to sufficient accuracy and the conditions under which the measurements were taken must be known.}
\item{Through the past century measurement of constants have become more precise and have changed. Examples of constants used in determining quoted magnetization and gyromagnetic data in the literature are the density of elements, the charge to mass ratio of the electron, and the Bohr magneton. Different groups use different values. Sometimes the values of constants used in calculations (eg. the Bohr magneton) are assumed to be known and are not given. }
\item{Experiments measuring properties of these ferromagnetic elements used different levels of purity. It is not clear what uncertainty should be assigned to account for the effects of impurities.}
\item{In many publications, the data are only shown as plots and the values of the measurements are not provided. The values must be extracted with plot digitization software. }
\item{In order to compare magnetization data taken with different sample shapes, the applied field must be converted to the internal field, $H_{\rm int}$. This conversion is not always possible if the data are not given in terms of $H_{\rm int}$ or the sample shape and dimensions are not provided so that this conversion from applied to internal field can be made.}
\end{itemize}

\

\subsection{Determining Saturation Magnetization}\label{method}
Target polarization is determined from measurements of the saturation magnetization. Another term used in the literature is ``spontaneous magnetization,'' which, as the name implies, refers to the magnetic moment of a material that spontaneously arises with no applied field. In ferromagnetic materials the magnetic moments of the electrons tend to spontaneously align in a given direction. However, due to energy considerations, domains tend to form in such a way that the total spin averaged across many domains at the macroscopic level is far below the saturation level and may be zero. In the presence of an applied magnetic field, the domain boundaries shift with enlarging domains having magnetic moments aligned along the direction of the field. As the applied field is increased, eventually the material will reach magnetic saturation where all the spins are aligned along the direction of the applied field. Thus, the saturation magnetization and the spontaneous magnetization are related quantities and spontaneous magnetization is numerically equal to the saturation magnetization at 0~K. Quoting from \cite{Kraftmakher2005}: ``Under a sufficiently high external magnetic field, the sample reaches saturation and represents a single-domain system oriented along this field direction. Therefore, the saturation magnetization can be considered to be equal (to) the spontaneous magnetization of one domain." For a discussion of domain formation and saturation magnetization see Kittel's Review paper from 1949\cite{KittelOct1949}.
\subsubsection{Temperature and Field Dependence of Saturation Magnetization \label{sec:sat_mag_THdep}}
Spontaneous magnetization is a function of temperature and applied field and for this reason it is often given as $M_{0}$, the value of saturation magnetization extrapolated to zero applied field at T = 0~K. However, experiments measure the magnetization at temperatures above 0 K with non-zero applied fields. For temperatures well below the Curie temperature and low applied fields, the magnetization has been shown to roughly follow the $T^{3/2}$ law of Bloch given as \cite{Bloch1930}
\begin{equation}
M_s(T) = M_0(1-a_{3/2}T^{3/2}),
\label{eq:bloch}
\end{equation}
where $M_0$ is the saturation magnetization at 0 K and $a_{3/2}$ is an empirically determined constant.  

This temperature-dependence of the saturation magnetization arises primarily from the presence of spin-waves which are traveling excitations of spin precessions about the magnetic field propagating through a material. Spin waves propagate via coupling between neighboring spins and are strongly temperature-dependent with thermal energy driving the excitations. Near absolute zero, spin waves are nearly absent and their increased effect with temperature causes saturation magnetization to decrease with temperature as the overall alignment of individual atomic moments with the applied field decreases. Increasing the applied field also decreases the effect of spin waves so that at  high fields and low temperature their effect is diminished. For a more detailed discussion of spin waves see \cite{Kittel1951,Dyson1956_2,Foner1969,PauthenetMar1982}.

At higher fields and temperatures not small compared to the Curie temperature additional terms are required beyond those included in Eq. \ref{eq:bloch}. Freeman Dyson used an expansion in powers of $T$ to parameterize the dependence of saturation magnetization on temperature and applied field\cite{Dyson1956,Dyson1956_2}. Frederic Keffer building on the work of Dyson and others developed a more elaborate form of the expansion with terms depending on $T^{3/2}$, $T^{5/2}$, $T^{7/2}$ and $T^2$ as well as the strength of the internal field\cite{Keffer1966}. The half-power terms in $T$ arise from spin waves and the $T^2$ term accounts for the possibility of Stoner-type excitations from the band structure in metals\cite{PauthenetNov1982}.

This parameterization, while accounting for temperature and field dependence arising from spin waves, fails to account for the nearly linear high-field paramagnetic susceptibility of ferromagnets well above saturation as well as effects unique to each sample which prevent saturation and thought to arise from impurities,  strains, anisotropy, domains and even the geometry of the sample\cite{Foner1969}. Foner {\it et al.} divide magnetization data into three regions: 1. the low-field region approaching saturation where the aforementioned sample-dependent effects prevent saturation at the theoretical saturation value and create curvature unique to each sample in the $M$ versus $H_{\rm{int}}$ curves just below saturation; 2. the high-field region above saturation where effects from spin waves and possible remnant anisotropy remain in addition to the high-field susceptibility; 3. and the ultra-high field region where magnetic phase transitions may exist and which is not of interest here\cite{Foner1969}. These considerations suggest that use of Keffer's parameterization may require additional terms to account for the linear high-field susceptibility as well as non-linear curvature in the approach to saturation. 

Pauthenet performed an extremely precise measurement of the saturation magnetization of Fe and Ni as a function of both temperature and internal field from 0 to 17~T. Pauthenet claims the absolute scale in his measurements is known only to $\pm$0.5\% due to uncertainty in calibration but that relative uncertainty is at the 0.01\% level, making his work an authoritative reference for high field corrections. Following the work of Keffer, he expressed the saturation magnetization $M$ as a function of temperature and internal field, while adding a term linear in applied field, $\chi(T)$, to account for the known effect of high field susceptibility:\cite{Keffer1966,PauthenetMar1982,PauthenetNov1982}
\begin{equation}
M(H_{\rm int}, T)=M_0\left(1-\sum_{s=\frac{3}{2},\frac{5}{2},\frac{7}{2}}a_s\frac{F(s,t_H)}{\xi(s)}T^s-a_2T^2\right) +\chi(T)H_{\rm int}.
\label{eq:pauthenet}
\end{equation}
Here $M_0$ is the spontaneous magnetization at 0~K and zero applied field, $F(s,t_H)=\sum_{p=1}^\infty p^{-s}e^{-pt_H}$ is the Bose-Einstein integral function, and $t_H=g\mu_BH_{\rm int}/k_BT$, where $g$ is the Land\'e g-factor, $\mu_B$ is the Bohr magneton, and $k_B$ is the Boltzmann constant. $H_{\rm int}$ is the internal field and  $\xi(s)$ is the Riemann zeta function. Pauthenet fits this parameterizaiton to his data to give numerical values for the coefficients, providing magnetization as a function of internal magnetic field and temperature (see Eq. 9, 10 and Table 1 from \cite{PauthenetMar1982}). We use Pauthenet's numerical parameterization of magnetization as a function of internal field and temperature provided in Eqs. 9 and 10 of \cite{PauthenetMar1982}, to make corrections for differences in temperature and internal field.

It is important to note the difference between internal field and applied field. In a manner somewhat analogous to the internal electric field cancelation inside a dielectric, the applied magnetic field is partially cancelled inside a ferromagnetic sample by its magnetization. The relationship between the internal field and the applied field is given by the following equation (in the cgs system)
\begin{equation}
H=H_{\rm int}+\frac{4\pi M}{\rho},
\label{eq:Hint}
\end{equation}
where $H$ is the applied field, $H_{\rm int}$ is the internal field, $M$ is the magnetization and $\rho$ is a demagnetization constant that depends on the shape of the sample. Since the internal field is thus partially cancelled by the magnetization, $4\pi M$ is sometimes referred to as the ``demagnetizing field''. 

Well below saturation, the internal field is nearly 0 due to the demagnetizing field. In the literature, field-dependent corrections are often given as a function of internal field $H_{\rm int}$ not applied field $H$. Above saturation magnetization, $H_{\rm int}$ is less than $H$ by the saturation magnetization (21.58 kOe for iron and 6.2 kOe for nickel). There appear to be errors in the literature that stem from incorrect exchanges of applied field and internal field. For example, Eq. 3 from deBever {\it et al.} incorrectly interprets Pauthenet's corrections as a function of flux density $B$ instead of internal field. As a result, they calculate a correction from an applied field of 1~T to the final value of 4~T. A 4~T field applied normal to a thin Fe foil such as they were discussing translates into an internal field of $\sim$1.8~T for Fe foils, requiring a smaller correction. C. D. Graham also appears to confuse the two in Fig. 5  of \cite{Graham1982} where he plots magnetization versus $1/H$ but combines data from multiple sources some of which are in terms of $1/H$ and others which are in terms of $1/H_{\rm int}$.  

\subsubsection{\label{sec:other_factors}Other Factors Affecting Magnetization Measurements}
There are several issues to be aware of when trying to interpret magnetization values quoted in the literature.

\begin{figure}
\centering
\includegraphics[width=0.7\textwidth]{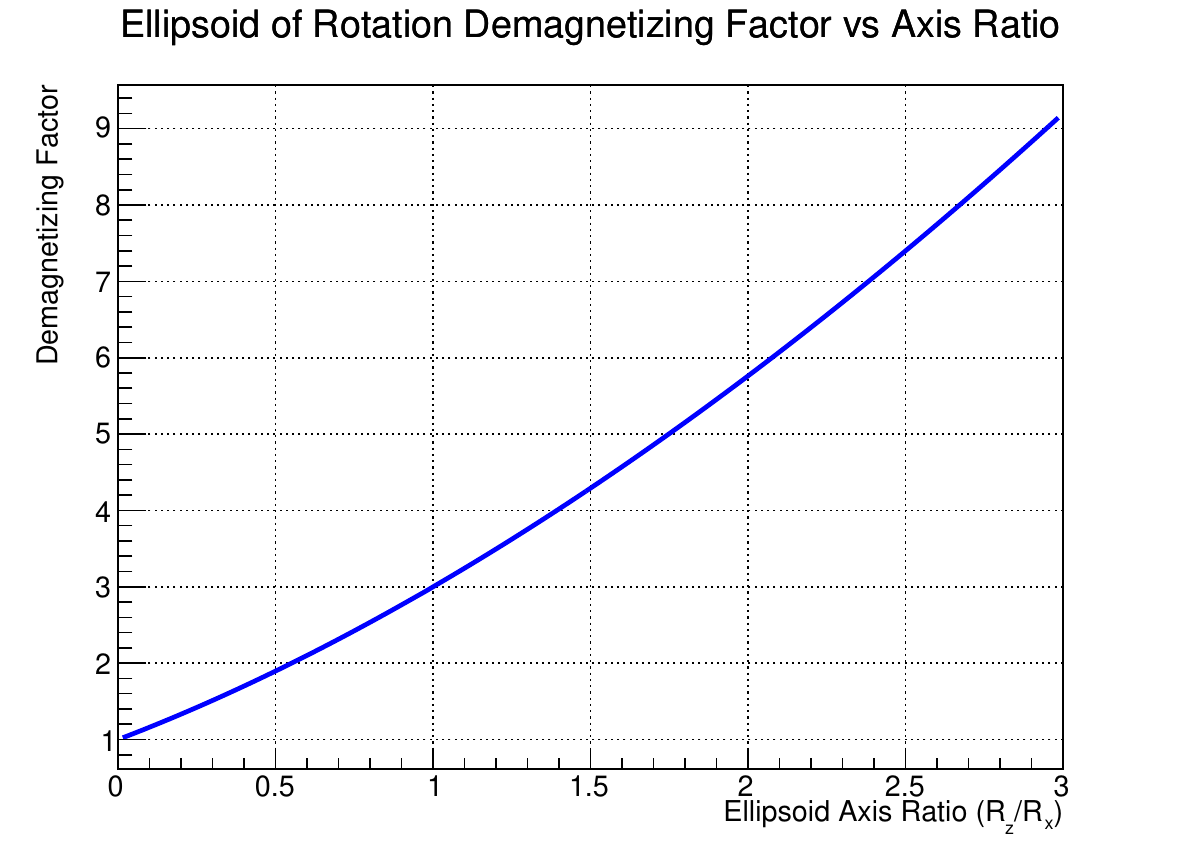}
\caption{Demagnetizing factor for ellipoids of rotation as a function of axis ratio for external magnetic field applied along the axis of rotation $R_z$. This plot uses equations 1a and 1b from \cite{Skomski2007}.}
\label{fig:demag_ellipsoid}
\end{figure}

{\bf Shape anisotropy:} the magnetization depends upon the shape of the object. Needles are very easy to magnetize along their long axis but much more difficult along a direction perpendicular to it. Each shape has a characteristic demagnetizing factor $\rho$ (see Eq. \ref{eq:Hint}) that is a function of the direction of applied field (unless symmetry dictates otherwise). Perfect spheres have a demagnetizing factor of 3. The demagnetizing factor for ellipsoids of rotation is a function of the ratio of the two axis lengths. Figure \ref{fig:demag_ellipsoid} shows the demagnetizing factor of ellipsoids of rotation as a function of the axis ratio where the applied magnetic field is along the axis $R_z$. A thin foil disk such as that used in the M\o ller polarimeter can be taken to be a flattened ellipsoid with an axis ratio of $\sim$0. In this case the demagnetizing factor approaches unity\cite{Skomski2007}.

{\bf Crystal anisotropy:} the crystal structure of a material can create directions along which it is easier to magnetize. The direction along which magnetic saturation is reached with the smallest applied field is called the easy axis of the crystal. Monocrystalline nickel, for example, has three different magnetization axes termed the [111], [110] and [100] axes, using standard Miller index notation, with [111] being the easy axis. Therefore, if one is using monocrystalline materials, the magnitude of the external field required to reach saturation will depend upon alignment of the crystal relative to the field. For polycrystalline materials there will be no preferred direction as a result of the random crystal orientations.

{\bf Crystal structure and phase changes}: some crystals have more than one possible crystal structure with different magnetizations. Their history of heating/cooling and annealing can have an effect on their magnetic properties. Cobalt, for example, goes through a phase change when heated at 690~K going from a close-packed hexagonal to a face-centered cubic crystal structure above 690 K which is unstable below that temperature. However, the exact crystal structure below 690 K (and by extension the magnetization) depends upon the grain size and the annealing process used to prepare it\cite{Owen1954}.

{\bf Stesses and strains:} stresses and strains in the material as well as porosity will affect how easily the material is magnetized. This can be seen particularly well by annealing, which often makes the material more easily magnetized\cite{Case1966}.

 \subsubsection{Measurements of Saturation Magnetization}
Although different methods are used to measure the saturation magnetization, they broadly break down into two categories: 
\begin{enumerate}
\item{ Force method: a small ellipsoid sample of the element of interest is placed in a precisely determined field gradient. With a proper setup, the force on the sample by the magnetic field can be shown to be the product of the magnetic moment of the sample and the magnetic field gradient. Thus the magnetic moment of the sample is given as the force divided by the field gradient. Dividing by the mass of the sample gives the mass magnetization directly. A possible source of systematic error in this method is the use of standard weights and a balance to measure forces. Conversion from  mass to force requires knowing the gravitational acceleration at the measurement location and relative uncertainty in this value translates directly into the final result. Of the magnetization measurements  included in this study, only those by Crangle {\it et al.} utilized this method. }
\item{ Induction method: a sample is placed into a magnetic field and its presence creates a magnetic moment that is measured in pickup coils. This directly measures volume magnetization and must be converted to mass magnetization by multiplying by density, introducing another potential source of systematic error.}
\end{enumerate}
     
Although the experimental methods can be thus broadly categorized, each individual experiment takes a slightly different approach to measurement and calibration.

Measurements of magnetization are performed at a variety of applied magnetic fields and temperature and are typically expressed in terms of the saturation magnetization $M_0$ which is the extrapolation to zero applied field at 0~K\cite{Crangle1971}. A review of the literature yields many measurements of the magnetization of iron and nickel. Different approaches can be taken to obtain ``consensus'' values. One approach taken by H. Danan {\it et al.}\cite{Danan1968} and deBever {\it et al.} \cite{deBever1997} is to average the values of spontaneous magnetization $M_0(H=0, T=0~K)$ and then apply a correction to obtain the magnetization at room temperature and nonzero applied fields. However, the process of extrapolation to zero field and temperature is not standardized and different methods are utilized, making this a poor standard for comparison. Furthermore, since we are looking for magnetization near room temperature this method introduces error extrapolating down to $M_0$ and once again correcting back up to room temperature and high fields. Since most measurements at least include data at or near room temperature and at internal fields at or close to 10~kOe (1~T), it makes sense to utilize magnetization measurements taken near room temperature and internal fields of order 10~kOe. Where the available data in the literature were not available at precisely T=294~K, small corrections were applied to the measurements based upon the formulation given in \cite{PauthenetMar1982}. In each case the data of magnetization versus internal magnetic field were parameterized using Eqs. 9 and 10 from \cite{PauthenetMar1982}.

Although the ``consensus'' values presented here for magnetization include data from a number of measurements done over a period from 1929-2001, this is not an exhaustive data set by any means. Table \ref{tab:magnetization_pubs} lists the publications used in this analysis for iron and nickel. We established the following criteria to decide which data to include:
\begin{itemize}
\item{Original data was published and publication was available. Some measurements referred to in the literature are not readily available. For example much of Danan's reported measurements on Ni were never published except in his 1968 review which provides few details of the experiment. }
\item{Data in the publication were available near room temperature $(294\pm10~\rm{K})$ and an internal field of 10 kOe. We corrected all data in this analysis to $T=294~$K. Starting with measurements of the magnetization close to these values of temperature and internal field keeps the corrections and extrapolation uncertainty small.}
\item{Enough details were provided to obtain the internal field of the sample either because the data were given versus internal field or the demagnetizing factor could be calculated from information given.}
\item{Data were taken with a high purity sample. With the exception of the NASA study by Behrendt {\it et al.} for which purity was not stated, all samples used had greater purity than 99.9\% to keep the systematic error from this source small. The NASA study was included in spite of the lack of information on sample purity because they claimed measurement error of $\pm$0.2\% and they were only the second data set we found with measurements in the high-field (several tesla) region of interest to us and which met the other criteria.}
\item{Systematic errors were sufficiently small to provide useful additional information. For example, Pauthenet \cite{PauthenetMar1982} has very precise data, but since he uses Danan's Ni data for absolute calibration, his systematic error is 0.5\%. Therefore, Pauthenet's data are used for relative corrections of field and temperature, but not in the absolute measurement average. Aldred \cite{Aldred1975} also has a precise data set, but calibrates his data using the ``known magnetization of nickel'' which is exactly what this analysis is seeking to determine. For this reason, we also did not retain Aldred's data.} 
\end{itemize}
\begin{table}[h]
\caption{\label{tab:magnetization_pubs}Publications used in obtaining consensus value for magnetization near room temperature at high fields.}
\begin{center}
\begin{tabular}{|l|l|l|l|}\hline
Publication & Year & T (K) & Comment\\\hline
Weiss and Forrer \cite{Weiss1929} & 1929 & 288 & Only Fe data used\\
R. Sanford {\it et al.}(NIST)\cite{Sanford1941} & 1941 & 298 & Data on Fe only\\
H. Danan \cite{Danan1959} & 1959 & 288 & Data on Ni and Fe\\
Arajs and Dunmyre \cite{Arajs1967}& 1967 & 298 & Data on Ni and Fe\\
Crangle and Goodman \cite{Crangle1971} & 1971 & 293 & Data on Ni and Fe\\
Behrendt and Hegland (NASA)\cite{Behrendt1972} & 1972 & 298.9 & Data on Fe only\\
R. Shull {\it et al.}(NIST) & 2000 & 298 & Data on Ni only\\\hline
\end{tabular}
\end{center}
\end{table}
\begin{figure}[ht!]
\centering
\includegraphics[width=0.9\textwidth]{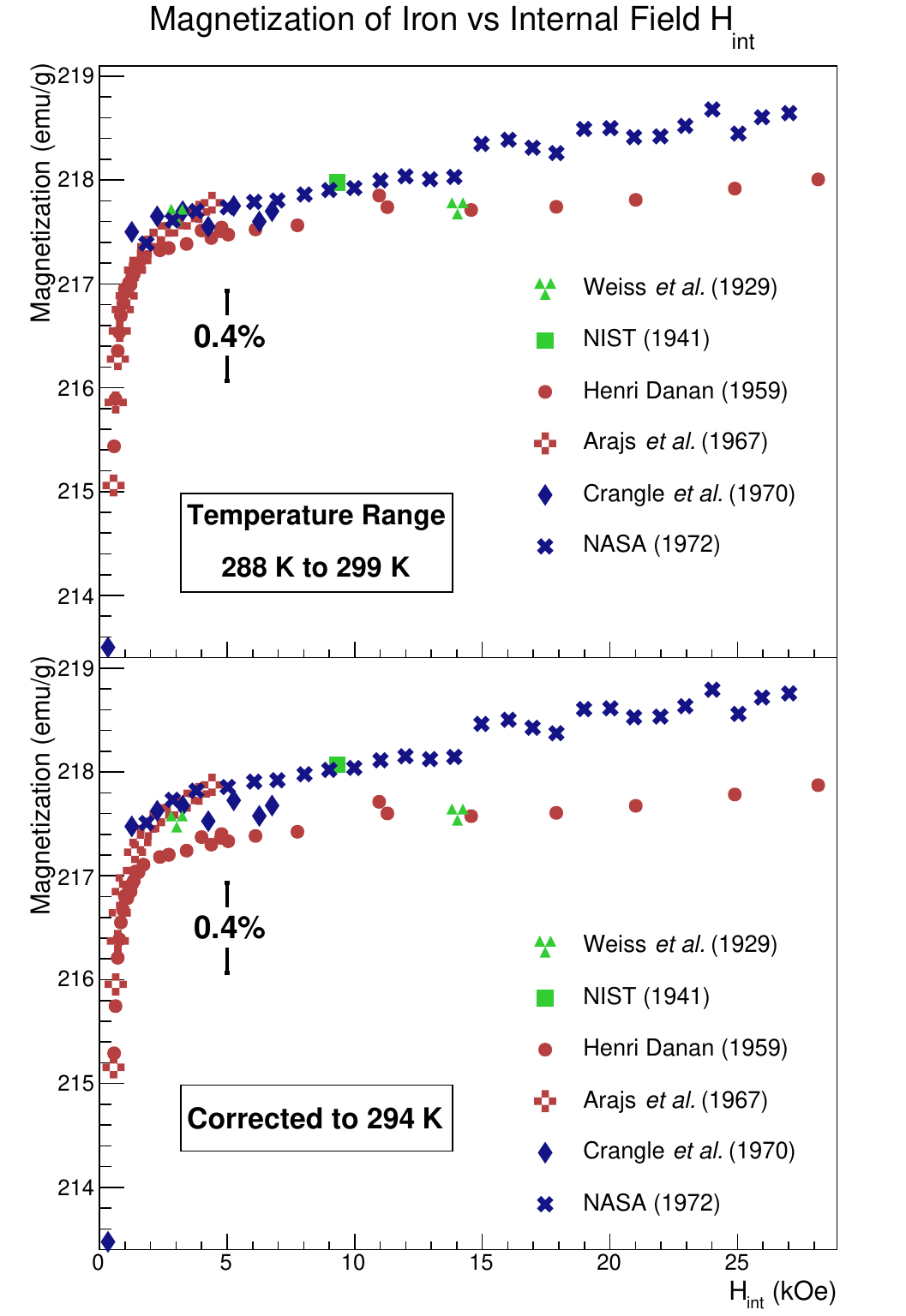}
\caption{Published magnetization data from various sources for Fe shown versus internal field. The top plot shows the data for the temperature at which it was taken and the the bottom plot shows the same data corrected to 294~K. Note that zero is suppressed on the vertical axis. Refer to Table \ref{tab:magnetization_pubs} for details on the data sets.}
\label{fig:mag_Fe}
\end{figure}
Fig.\ref{fig:mag_Fe} shows the data for the magnetization of Fe from the published sources before and after correction to $T=294$~K. Where data were not given in terms of internal field $H_{\rm int}$, they were converted to $H_{\rm int}$ using Eq. \ref{eq:Hint} using information given in the publications to determine the demagnetizing field $4\pi M/\rho$. The data are approximately linear as expected in the high-field region above 3~kOe. The lower panel of Fig. \ref{fig:mag_Fe}  shows the data after correction to the standard temperature 294~K. It is striking that the temperature correction increases the inconsistency between the different data sets. As previously mentioned, the temperature correction was taken from Pauthenet's parameterization given in Eq 9 in \cite{PauthenetMar1982} (see Eq. \ref{eq:pauthenet}) with the coefficients found empirically to be $a_{3/2}=307\times 10^{-6}$,  $a_{5/2}=-22.8\times10^{-8}$ and $a_{7/2}=0$. Pauthenet evaluates the factor $g\mu_B/k_B$ as $1.378\times10^{-4}$.\footnote{Note that Pauthenet actually gives $g\mu_B/k_B=1.378$ for Fe in Eq. 9  of \cite{PauthenetMar1982}, but  replicating his plots in Figure 1 of \cite{PauthenetMar1982} requires an extra factor of $10^{-4}$.} A linear approximation $\chi(T)=3.644\times10^{-6}+5.0434\times10^{-10}T$ was obtained from a fit to the discrete data points provided in Table 1 of \cite{PauthenetMar1982} in order to be able to evaluate $\chi(T)$ for any temperature. 

\begin{figure}
\centering
\includegraphics[width=0.9\textwidth]{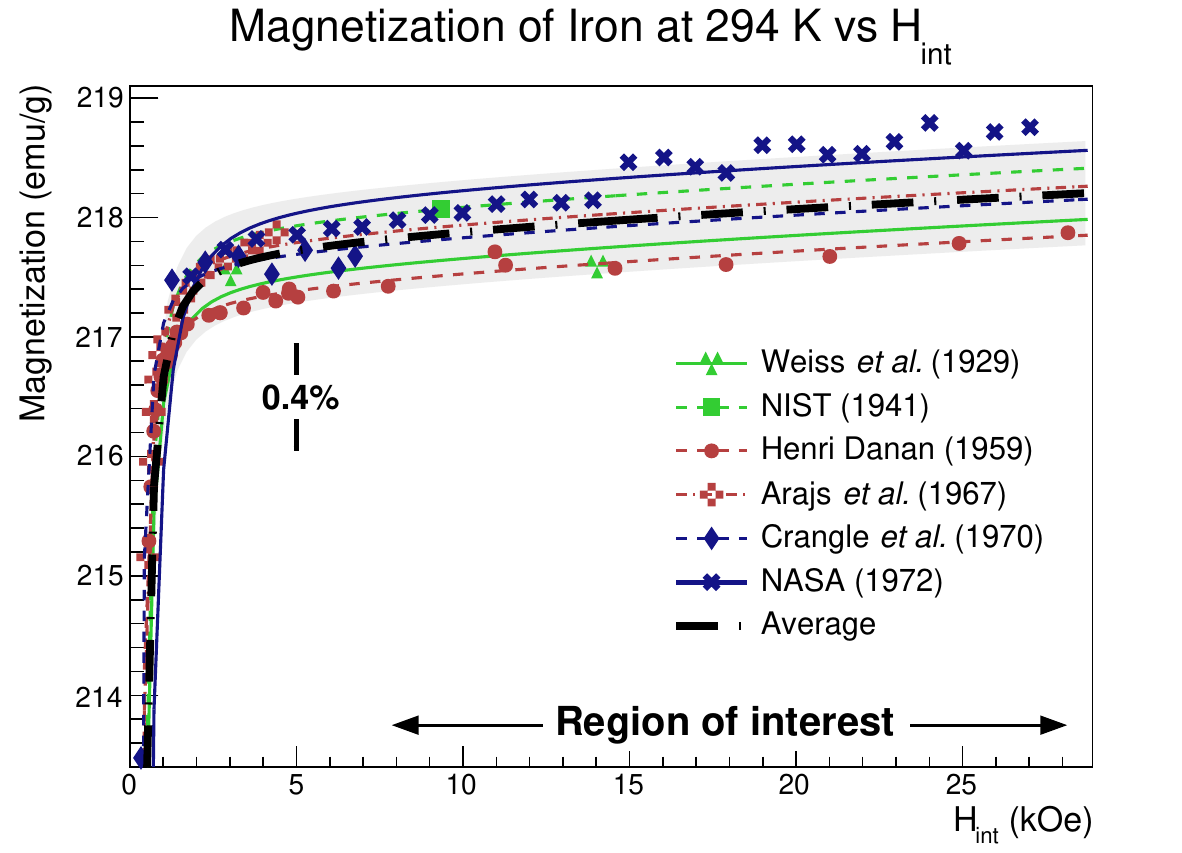}
\caption{Published magnetization data from various sources for Fe plotted versus internal field corrected to 294~K. Magnetization data are fit using a modified form of Eq. 9 from \cite{PauthenetMar1982}. Each of the six datasets are fit individually and the resulting curve fits averaged (see text for details). The error band corresponds to $\pm$0.20\% or $\sim$0.44 {\rm emu/g}. }
\label{fig:mag_errorband_Fe}
\end{figure}

To get an average parameterization versus internal field, each of the six temperature-corrected data sets were fit individually using Pauthenet's parameterization with $T=294~K$ as can be seen in Fig. \ref{fig:mag_errorband_Fe}. Pauthenet's work was chosen as the high-field reference since he quotes the relative uncertainty of the data used in his fit to be at the 0.01\% level and his parametrization in the high-field region accurately reproduces the field dependence seen in the data. 

An additional term of $a/H_{\rm int}^2$ was added to Pauthenet's parameterization to provide a better fit at low internal field in the approach to saturation. Pauthenet's data  did not roll off as quickly as the data used here (see Fig. 1 of \cite{PauthenetMar1982}). The exact curvature in this region is expected to depend on the composition and purity in addition to stresses and imperfections in the sample used which will vary from sample to sample. Pauthenet used a high purity monocrystalline sample aligned along the easy axis to suppress effects from anisotropy and strains, whereas many of the datasets included here used polycrystalline samples, providing a plausible explanation of the discrepancies in this region. 

Stoner discusses the interpretation of terms proportional to $1/H_{\rm int}$ as arising from inclusions (impurities or cavities) in the sample and $1/H_{\rm int}^2$ as arising from stresses and imperfections (see discussion around Eqs. 4.18-4.22 in \cite{Stoner1950} and around Eq. 7 of \cite{Foner1956}). 

For the Fe datasets included here, the term proportional to $1/H_{\rm int}$ was not needed, so only a term of the form $a/H_{\rm int}^2$ was retained. The coefficient $a$ was constrained to values 0 or below in the fit to maintain consistency with the physics model. For the data sets with measurements over a range of $H_{\rm int}$ both $M_0$ and $a$ were used as fit parameters. In fits for two of the data sets (Weiss {\it et al.} and Sanford {\it et al.}), only $M_0$ was allowed to float due to the limited number of data points and $a$ was fixed to the average from the data sets where it was allowed to float as a fit parameter. The data for Weiss and Forrer were not specifically given, but the following linear parameterization was provided from a fit to data over the range of applied fields from 0.6 to 1.7~T: \cite{Weiss1929}
\[
M_0(H)=217.76\left(1-\frac{2.6}{H}\right),
\]
where $H$ is the applied field in oersteds. This parameterization was used to determine two data points at 0.6~T and 1.7~T which were then fit to determine $M_0$. The data for Sanford (NIST) {\it et al.} are condensed in the literature to a single value of $H_{\rm int}$ even hough they are composed of multiple values across a range of applied fields not included in the publication. 

The average value of $M_0$ and $a$ from the fits were used to produce the average parameterization curve shown. Over the range of $H_{\rm int}$ from 8 to 28~kOe (about 3 to 5~T applied field for a thin Fe foil magnetized out of plane normal to the surface) the following second degree polynomial accurately follows the average parametrization curve: 
\begin{equation}
\label{eq:param_MvsHint}
M_{\rm sat}^{\rm (Fe)}(H_{\rm int},294~K)=217.628+2.7439\times10^{-2}H_{\rm int}-2.6304\times10^{-4}H_{\rm int}^2,
\end{equation}
where $H_{\rm int}$ is in units of kOe. This parameterization is shown in Fig.\ref{fig:mag_errorband_Fe}. A systematic error band of $\pm$0.20\% is assigned to account for the spread of the data. The source of this systematic spread across the datasets is not clear. 

Using 2.157~T for the magnetic saturation induction $(4\pi M_{\rm sat})$ of iron and a demagnetizing factor of unity for a thin foil magnetized out of plane, gives an internal field which is 2.157~T less than the applied field near saturation. Thus a uniform external 4~T magnetic field corresponds to an internal field of approximately 1.84~T. Converting Eq. \ref{eq:param_MvsHint} to applied field $B_{\rm app}$ in Tesla (this is the field of the magnet alone without the induction of the foil) for the specific case of a thin foil magnetized out of plane gives the following second order polynomial parameterization accurate over the region of 3-5~T applied field:
\begin{equation}
M_{\rm sat}^{\rm(Fe)}({\rm emu/g})=216.914+0.387863\hspace{.06667em}B_{\rm app}-0.026304\hspace{.06667em}B_{\rm app}^2.
\end{equation}
This gives the saturation magnetization per gram for iron at 294~K with an applied field of 4~T as $M_{\rm sat}^{\rm (Fe)}=218.04\pm0.44$~emu/g. This translates into $2.1803\pm0.0044~\mu_B/$atom which differs slightly from the value of $2.183\pm0.002~\mu_B/$atom determined by deBever {\it et al.}\cite{deBever1997} partially due to their over-correction for the magnetic field dependence. The small uncertainty quoted by  deBever {\it et al.} comes from C. D. Graham's review \cite{Graham1982} and uses the single data set of Crangle {\it et al.}\cite{Crangle1971} with a 0.1\% uncertainty. Furthermore, this publication by deBever {\it et al.} also misinterprets the 1~T applied field for Crangle's elliptical sample as being equivalent to a 1~T applied field for a thin foil magnetized out of plane.  While the data used in this analysis include that of Crangle {\it et al.} (see Fig. \ref{fig:mag_Fe}), we judge the uncertainty to be considerably greater than 0.1\% based on the spread in the various data sets.

\begin{figure}
\centering
\includegraphics[width=0.84\textwidth]{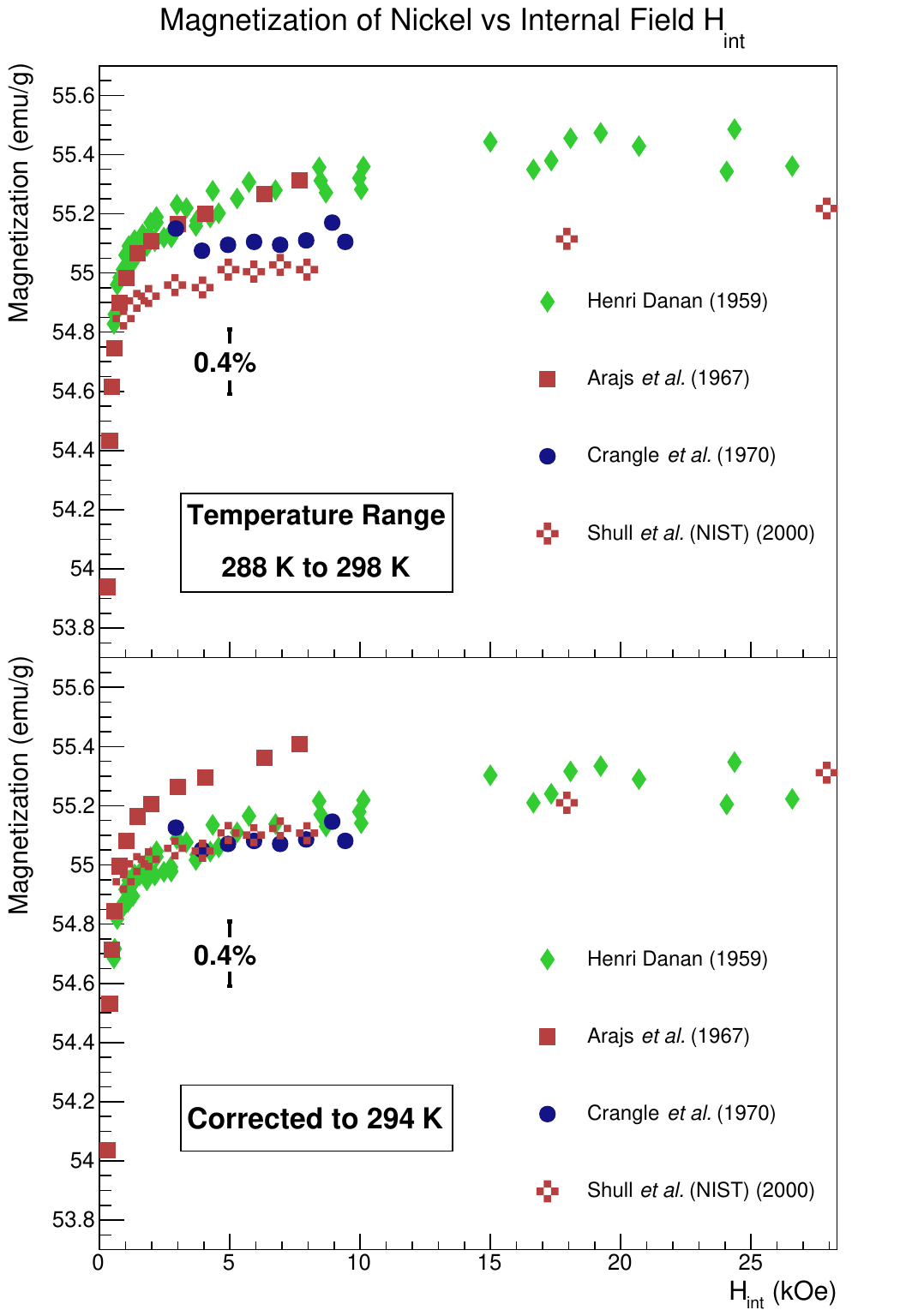}
\caption{Published magnetization data from various sources for Ni shown versus internal field. The top plot shows data for temperature at which it was taken and the bottom plot shows the same data corrected to 294~K. There is good agreement in the data with the clear exception of that from Arajs {\it et al.} which are systematically higher by $\sim0.5\%$. The reason for this discrepancy is not clear. Their publication claims $\pm$0.2\% accuracy for saturation magnetization which cannot explain the full difference.}
\label{fig:mag_Ni}
\end{figure}

\begin{figure}
\centering
\includegraphics[width=0.9\textwidth]{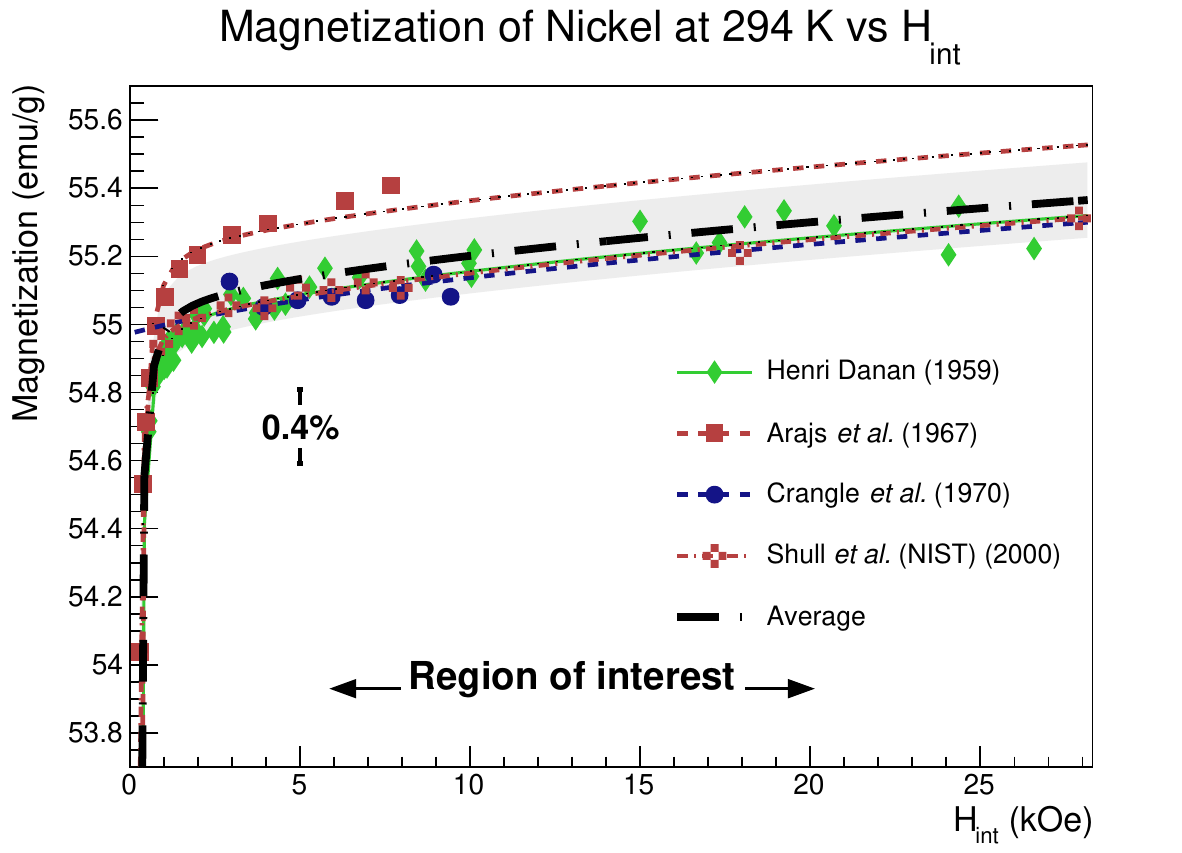}
\caption{Published magnetization data from various sources for Ni plotted versus internal field corrected to 294~K and shown with proposed parametrization curve for internal fields up to 20~kOe (2~T). Magnetization data are fit using a modified form of Eq. 9 from \cite{PauthenetMar1982}. Each of the six datasets are fit individually and the resulting curve fits averaged (see text for details). The error band corresponds to $\pm$0.20\% or $\sim$0.11~{\rm emu/g}.}
\label{fig:mag_errorband_Ni}
\end{figure}

A similar analysis of the literature for nickel is shown in Fig. \ref{fig:mag_Ni}. As for Fe, the Ni data were fit to the Pauthenet parameterization with an additional term of $a/H_{\rm int}^2$. Each of the four data sets were fit independently in $M_0$ and $a$ with $a$ being constrained to be 0 or less as before. The only exception to this parameterization was the Crangle data set where $a$ was fixed at 0 since there were no low field data to guide the fit. The fits are shown in Fig. \ref{fig:mag_errorband_Ni}. The ``Average" parameterization curve was formed using the average $M_0$ and $a$ from the fits. This average parameterization along with a proposed systematic error band of $\pm0.2$\% or 0.11~emu/g is shown in Fig.\ref{fig:mag_errorband_Ni}.  Using 0.6179~T for the magnetic saturation induction of nickel and a demagnetization factor of unity for a thin foil magnetized out of plane, makes the internal field 0.6179~T less than the applied field near saturation. Thus a uniform external 2~T magnetic field corresponds to an internal field of approximately 1.38~T. Over the range of $H_{\rm int}$ from 6 to 20~kOe (approximately 1.2 to 2.6~T applied field for a thin Ni foil magnetized out of plane normal to the surface) the following polynomial precisely follows the fit parameterization curve: 
\begin{equation}
M_{\rm sat}^{\rm (Ni)}({\rm emu/g})=55.063+1.5718\times 10^{-2}H_{\rm int}-1.9678\times 10^{-4}H_{\rm int}^2,
\label{eq:param_MvsHintNi}
\end{equation}
with $H_{\rm int}$ in units of kOe. Converting Eq. \ref{eq:param_MvsHintNi} to applied field $B_{\rm app}$ in Tesla for the specific case of a thin Ni foil magnetized out of plane:
\begin{equation}
M_{\rm sat}^{\rm (Ni)}({\rm emu/g})=54.959+0.181495\hspace{.06667em}B_{\rm app}-0.019678\hspace{.06667em}B_{\rm app}^2.
\end{equation}
This gives the magnetization per gram for nickel at 294~K with an applied field of 2~T as $M_{\rm sat}^{\rm (Ni)}=55.24\pm0.11$~emu/g. This translates into $0.5806\pm0.0012~\mu_B/$atom

\FloatBarrier
\subsubsection{Magnetocrystalline anisotropy}
As previously discussed in section \ref{sec:other_factors}, the crystal structure of ferromagnetic elements  creates axes along which it is easier or harder to magnetize the material. The origin of this anisoptropy is primarily from the spin-orbit coupling. The spin-spin coupling works to align adjacent spins in either parallel or anti-parallel orientations but does not couple to the crystal lattice. The spin-spin coupling can be rotated relatively easily with external magnetic fields. Conversely, the orbital magnetic moments are strongly coupled to the crystal lattice such that even very strong magnetic fields do not easily rotate them. The coupling between the spin and orbital motion of each electron tends to align the spins of the electrons along the crystal lattice such that there is an additional energy associated with rotating the spins away from what is termed the ``easy axis" of the crystal. This coupling is also relatively weak with fields of a few hundred oersteds being sufficient to overcome it. For a more detailed discussion refer to {\it An Introduction to Magnetic Materials} by Cullity and Graham section 7.4\cite{Cullity2008}.

\begin{figure}[ht]
\centering
\includegraphics[width=0.99\textwidth]{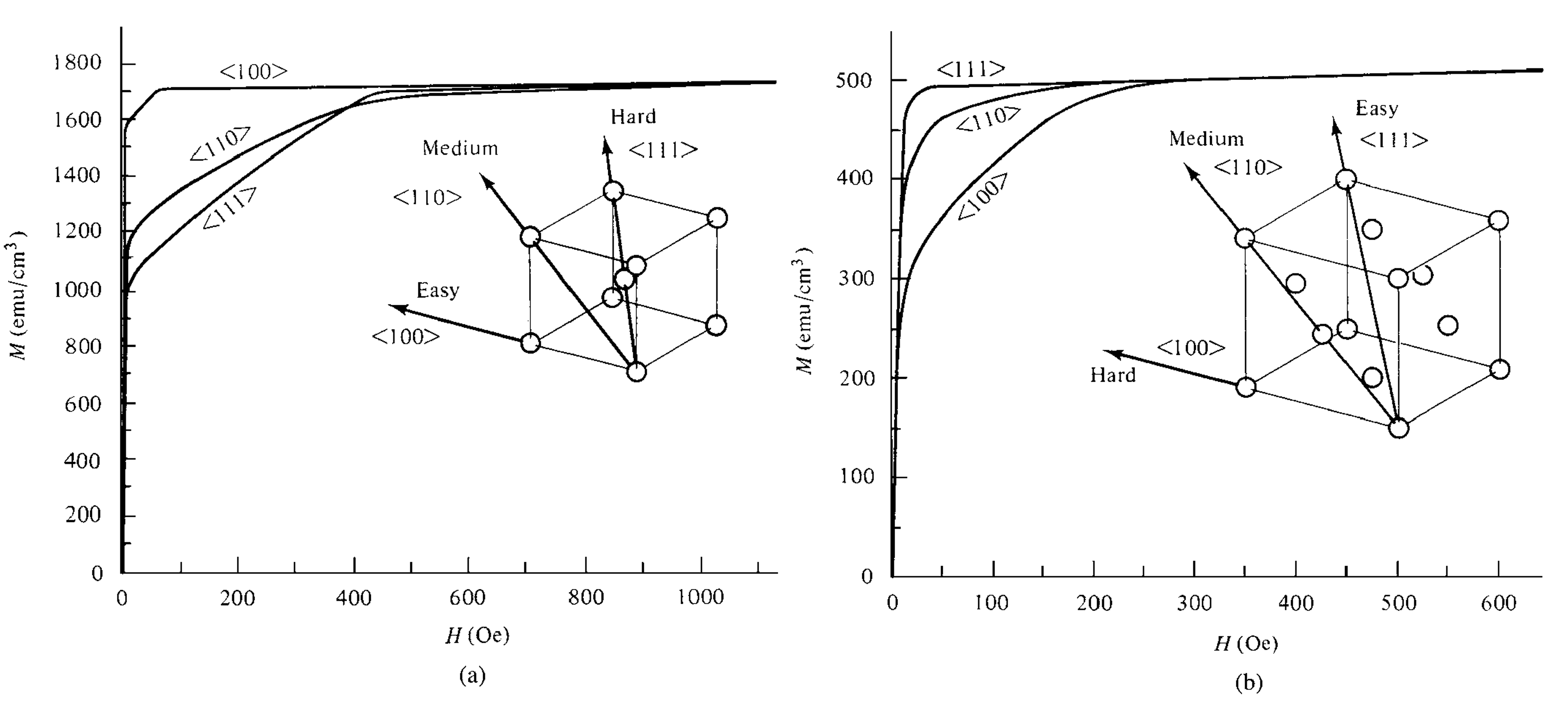}
\caption{Magnetization curves for single crystals of Fe (a) and Ni (b) demonstrating the relative difficulty of magnetizing the crystals along different directions. (Figure adapted from \cite{Cullity2008}.)}
\label{fig:anisotropy_Ni_Fe}
\end{figure}

Iron and nickel (iron is body-centered cubic and nickel is face-centered cubic) have hard, medium and easy magnetization axes due to their crystal lattice structure. Magnetization along any axis other than the easy axis requires a larger applied magnetic field due to the anisotropy energy. The plots in Fig. \ref{fig:anisotropy_Ni_Fe} show typical magnetization curves for iron and nickel along each of their magnetocrystalline axes. It is important to note that each of the magnetization curves in Fig. \ref{fig:anisotropy_Ni_Fe} appears to approach the same saturation magnetization. Pauthenet measured the saturation magnetization with precision along the different crystallographic axes for Ni and Fe and concluded that the saturation magnetization is the same to within 0.01\% at an internal field of 10 kOe or greater\cite{PauthenetNov1982}.
\subsubsection{Discussion of cobalt as a potential target material}
Two key features of cobalt make it unfit as a precision target material. First, the crystal structure of cobalt (mainly close-packed hexagonal at room temperature) creates a greater magnetocrystalline anisotropy than it does for the other two ferromagnetic elements. Pauthenet measured the difference in saturation magnetization along the different axes to be at the 0.5\% level in his careful study of magnetization versus field\cite{PauthenetNov1982}. In a polycrystalline sample such as a foil that might be utilized in the M\o ller polarimeter, it is not apparent how to determine the saturation magnetization. 

Second, the crystal structure of cobalt changes from primarily close-packed hexagonal below 690~K to face-centered cubic above this temperature. Near room temperature, a mixture of the two crystal structures generally of which the fractional composition varies from sample to sample producing a large uncertainty in the saturation magnetization for this material\cite{Myers1951}. For these reasons, we have discarded cobalt as a candidate precision target material.

\FloatBarrier
\newpage
\subsubsection{\label{sec:target_heating}Target heating and temperature corrections}
\begin{wrapfigure}{r}{0.3\textwidth}
\vspace{-20pt}
\centering
\includegraphics[width=0.25\textwidth]{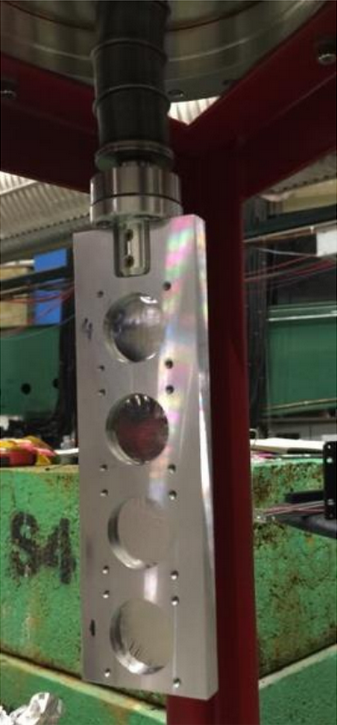}
\caption{Target ladder with four thin iron foil disks. The support structure is aluminum.}
\label{fig:target_ladder}
\end{wrapfigure}
The magnetization of Fe and Ni is found for room temperature; however, there is a relatively large temperature-dependent correction ($\sim$1.5\% from liquid helium to room temperature for Fe) to the saturation magnetization as discussed in section \ref{sec:sat_mag_THdep}. We now discuss the temperature corrections to the target magnetization for temperatures above 294 K that would be created by heating of the target by the electron beam. Note that although the following analysis is specific to the Hall A setup (circular foil, circular electron beam centered on the foil, un-rastered Gaussian profile electron beam). 
Further details of the calculation that allow it to be extended beyond these specific parameters can be found in \cite{JonesDC2022}.

When the electron beam is on target during a M\o ller polarimetry measurement, energy deposition causes the foil to heat up by a few degrees under usual conditions. Since there is a slight temperature dependence to the magnetization a correction will have to be applied. The further from the Curie temperature of the material, the smaller the correction will be. Therefore, we can expect the beam heating correction for Ni to be fractionally larger than that of Fe (see Table \ref{tab:ferro_elem}). 

In the absence of a direct way of determining the temperature of the foil at the beam spot during operation or of monitoring the relative magnetization {\it in situ}, an estimate of the temperature increase must be made. This section provides a calculation of the foil heating from the electron beam under a set of assumptions.

The thin foil circular disks used in the M\o ller polarimeter are a few microns thick (see Fig. \ref{fig:target_ladder}). The electron beam flux profile is approximately Gaussian with a typical 1$\sigma$ radius of 100~$\mu$m. 

The beam is approximately centered on the M\o ller target and has a natural helicity-correlated jitter of a few tens of microns. We calculate the approximate foil temperature change based on a few reasonable assumptions. We assume the beam introduces a heat load that is approximately a circular Gaussian distribution centered on the foil disk and that radiative black-body cooling is negligible. We also assume that the aluminum frame constitutes an approximately infinite heat sink i.e. the temperature of the aluminum frame remains at or near room temperature, and that the foils are 0.65~inch in diameter and in perfect thermal contact with the aluminum frame along their edges.

\begin{figure}[ht]
\centering
\includegraphics[width=0.7\textwidth]{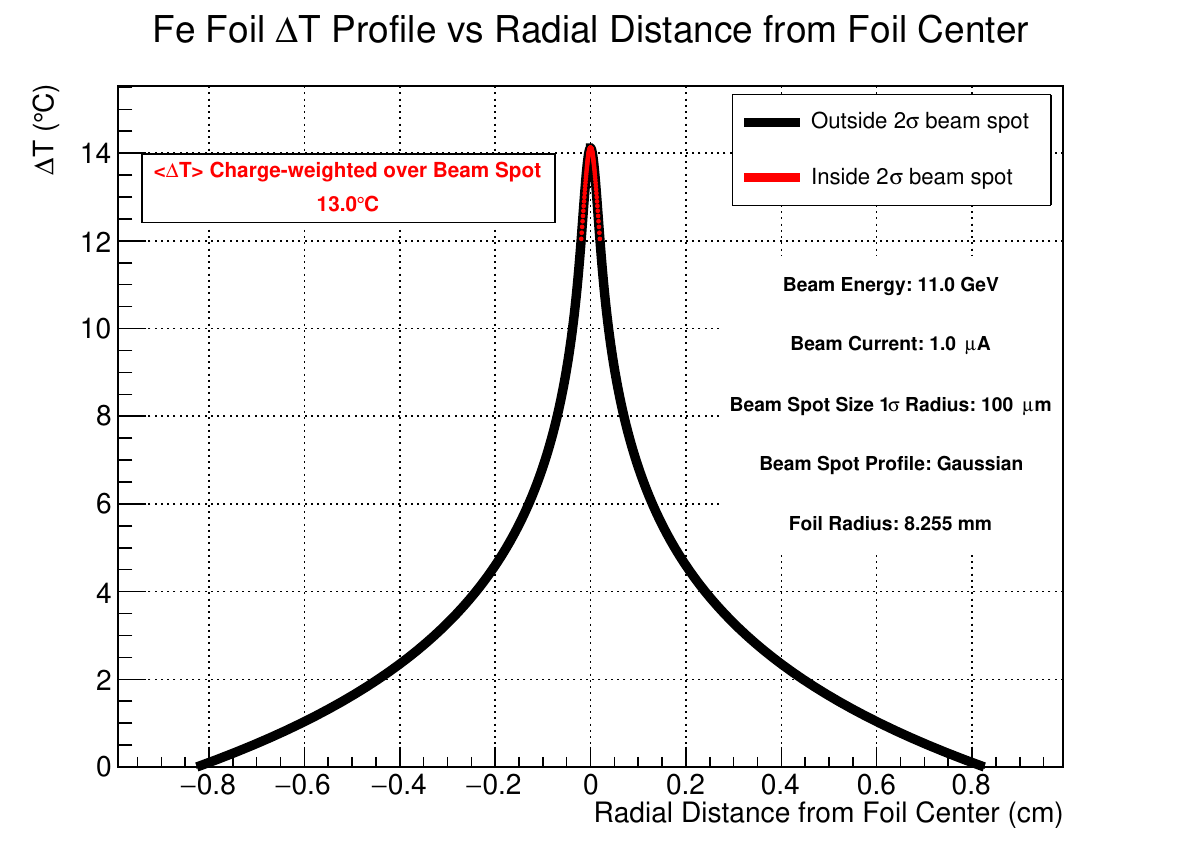}
\caption{Foil temperature distribution in a 0.65~inch diameter foil under a 1~$\mu$A beam load. The electron beam is assumed to have a Gaussian distribution with a beam current and energy, foil radius and 1$\sigma$ beam radius given in the plot. The red tip of the distribution is the part of the foil inside the 2$\sigma$ beam spot. The  average temperature rise weighted by the beam distribution over the beam spot is also shown. The {\sc Root} macro for making this plot is called ``FeFoilHeating.C" and is available at the following Git repository: https://github.com/jonesdc76/MollerPolarimetry/blob/master/TargetPolarization/}
\label{fig:target_heating}
\end{figure}

\begin{figure}[ht]
\centering
\includegraphics[width=0.7\textwidth]{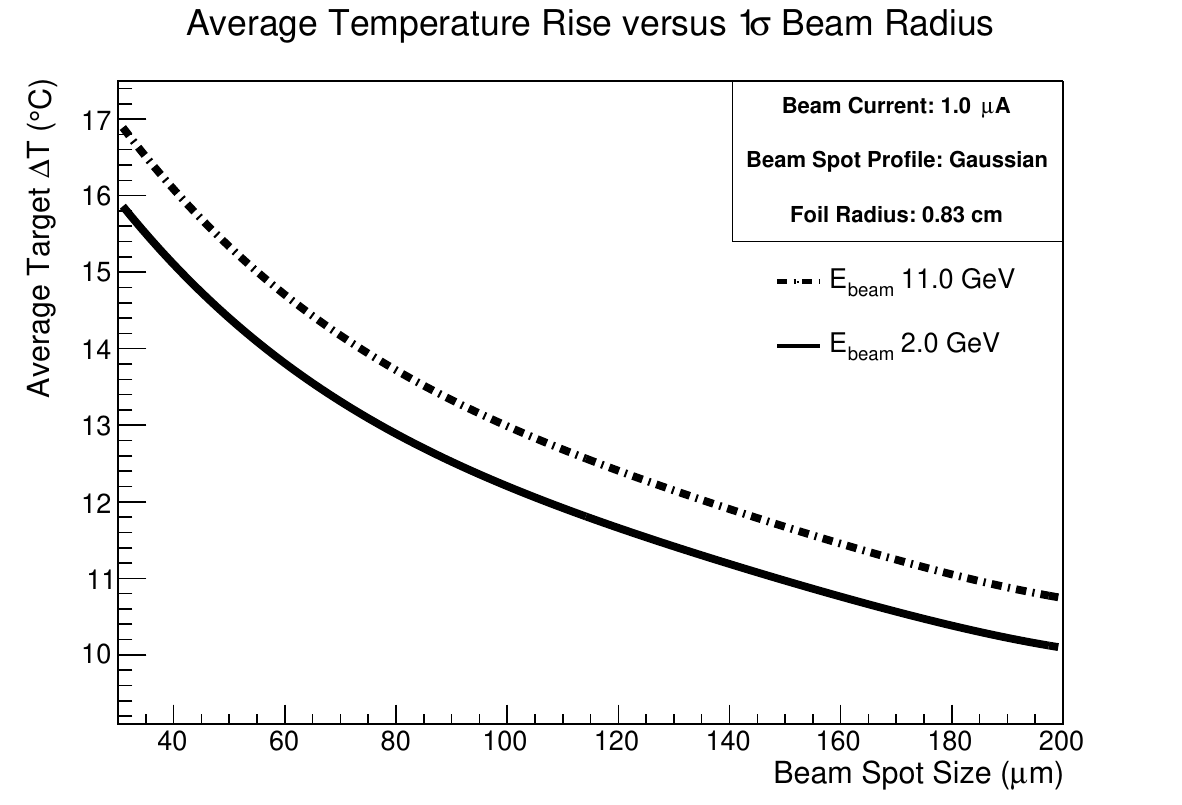}
\caption{Average foil temperature increase (weighted by the beam charge distribution) shown versus beam spot size radius for the parameters shown.}
\label{fig:spotsize_dep}
\end{figure}

The heat equation for this situation with only radial dependence and in the steady state is given as
\begin{equation}
\kappa\nabla^2T=-\rho\alpha B_{\rm flux},
\end{equation}
which reduces to
\begin{equation}
\label{eq:heat_T_r}
\frac{\partial}{\partial r}\left(r\frac{\partial T}{\partial r}\right)=-\frac{\rho\alpha}{\kappa}rB_{\rm flux},
\end{equation}
where $\kappa$ is the temperature dependent thermal conductivity of Fe; $\rho =7.874$~g/cm$^3$ is the density of Fe; $\alpha$ is the collision stopping power for electrons in Fe, which is a function of electron energy; and $B_{\rm flux}=\frac{d^3N_e}{ds dt} $ is the flux density of the beam in $e^-$/(cm$^2$ s). This equation can be easily solved numerically with a Gaussian beam profile $B_{\rm flux}$ proportional to $e^{-r^2/2r_b^2}$, where $r_{b}$ is the 1$\sigma$ radius of the beam. The solution is shown in Fig. \ref{fig:target_heating} with a 1~$\mu A$ beam heat load with a typical spot size of $r_{b}=100~\mu$m. Fig. \ref{fig:spotsize_dep} shows the dependence of the average temperature rise on the beam spot size for otherwise similar parameters.
Using these data we obtained a temperature rise of 13.0$^\circ$C/$\mu$A for Fe as shown in Fig. \ref{fig:target_heating}. A similar temperature rise of 13.2$^\circ$C/$\mu$A was found for Ni foil. An ANSYS-Fluent simulation of heating for Fe foils under similar assumptions was found to agree at the 0.1$^\circ$C with the temperature rise calculation detailed here or a 1~$\mu$A  heat load on a 10~$\mu$m thick foil. 

\begin{figure}[ht]
\centering
\includegraphics[width=0.7\textwidth]{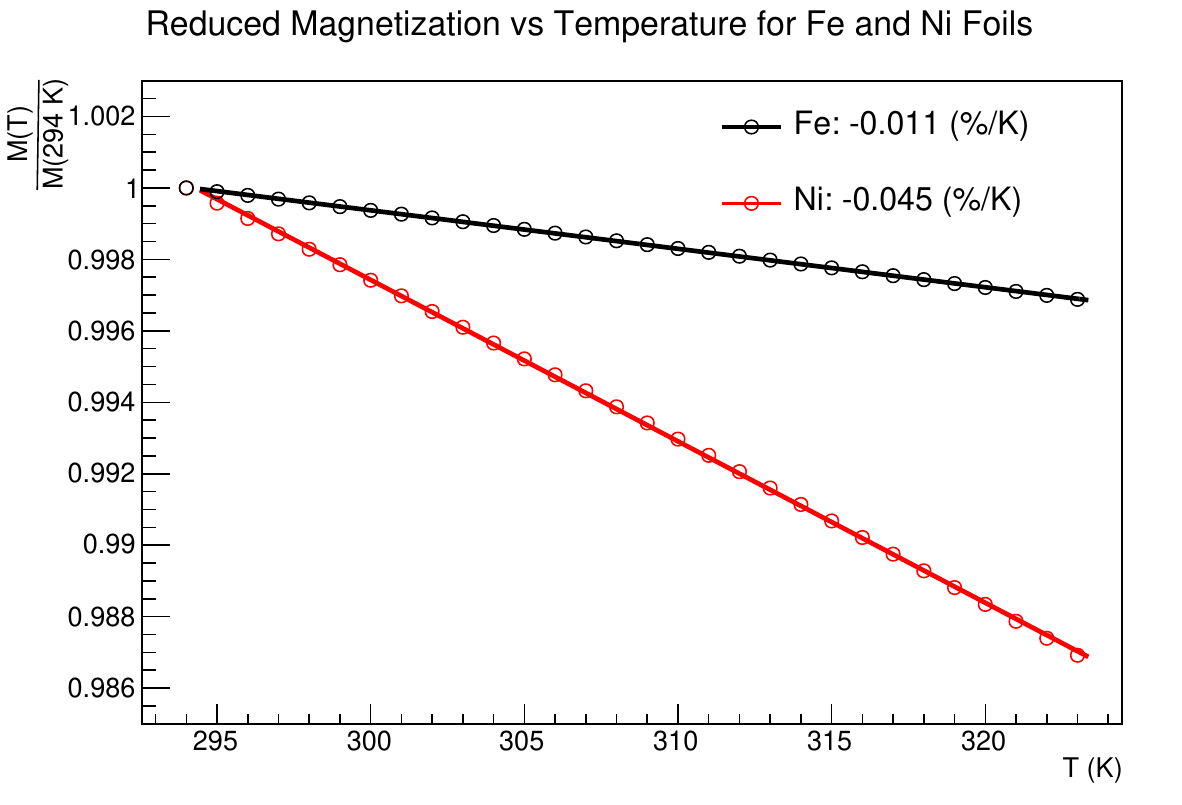}
\caption{Magnetization versus temperature as a fraction of its value at 294~K from the parameterization in \cite{PauthenetMar1982,PauthenetNov1982} and evaluated at an applied field of 2~T for an Ni foil and 4~T for Fe. The fractional temperature correction given by model is shown as a linear fit and is -0.011\%/K (-0.024~emu/g K) for Fe and -0.045\%/K (-0.025~emu/g K) for Ni.}
\label{fig:temperature_correction}
\end{figure}

The temperature dependence of magnetization for iron and nickel from \cite{PauthenetMar1982,PauthenetNov1982} yields the sensitivity shown in Fig. \ref{fig:temperature_correction}. The model was evaluated for applied fields of 2~T for nickel and 4~T for iron. A linear fit yields correction slopes of -0.025~(emu/g/$^{\circ}$C) for Ni and  -0.024~(emu/g$/^{\circ}$C) for Fe. A conservative uncertainty of 30\% is sufficient to cover both the uncertainties from the calculation of temperature increase and the magnetization versus temperature correction slope, yielding an uncertainty in the magnetization of $\pm$0.09~({\rm emu/g}/$\mu$A) for both Ni and Fe. 

\FloatBarrier

\subsubsection{Effect of impurities}
We next consider the effect of impurities on the measured magnetization. The experiments whose data are used in this analysis (with the possible exception of the measurement at NASA by Behrendt {\it et al.}) utilized highly pure Fe and Ni samples. Table \ref{tab:impurities} lists the level of impurities in the samples used in the various experiments whose data are used in this analysis. Although Weiss and Forrer \cite{Weiss1929} do not give a numerical value for the level of impurities they assure us that there were no impurities at a measurable level. They used this highly pure sample for the most precise results and many samples of less pure iron for less accurate studies. To set the scale, their less pure sample had a total of 0.22\% impurities with 0.09\% of that being carbon. Although the NASA measurement by Behrendt {\it et al.} does not list a purity level for the sample, we retain this measurement in spite of this uncertainty since it is only the second data set we found with precision measurements in the high field region (4~T applied fields) where we are typically running. An appropriately large systematic error is assigned in the end to account for this uncertainty.
\begin{table}[]
\caption{Level of impurities from the various measurements used in this analysis. Note that Danan used the same Fe sample measured by Weiss and Forrer. Crangle and Goodman used two samples for Fe and two for Ni of differing purities.}
\begin{center}
\begin{tabular}{|l|c|c|}\hline
Experiment&Element&Impurity Fraction\\\hline
Weiss and Forrer \cite{Weiss1929} & Fe&``No detectable impurities"\\
R. Sanford {\it et al.}(NIST)\cite{Sanford1941} & Fe&$<$0.01\%\\
H. Danan \cite{Danan1959, Danan1968}& Fe&Same as Weiss and Forrer\\
Arajs and Dunmyre \cite{Arajs1964}\cite{Arajs1967}&Fe& $\sim$600~ppm\\
Crangle and Goodman \cite{Crangle1971} & Fe & 0.06\% and 0.006\%\\
Behrendt and Hegland (NASA)\cite{Behrendt1972} & Fe & Not given\\\hline
H. Danan \cite{Danan1959, Danan1968} & Ni&0.01\% \\
Arajs and Dunmyre \cite{Arajs1963, Arajs1965, Arajs1967}&Ni& $\sim$30~ppm\\
Crangle and Goodman \cite{Crangle1971} & Ni & 0.05\% and 0.005\%\\
R. Shull {\it et al.}(NIST) & Ni & 10~ppm\\\hline
\end{tabular}
\end{center}
\label{tab:impurities}
\end{table}%

Addition of non-ferromagnetic impurities typically decreases the magnetization (see for example \cite{Luborsky1980, Ahern1958, Sanford1941}). Sanford {\it et al.} corrected for the effect of $\sim0.01\%$ impurities which yielded a correction at the $\sim0.02\%$ level\cite{Sanford1941}. Ahern {\it et al.} also found that adding copper to nickel reduced the magnetization by about 2\% for every 1\% of the nickel replaced by copper. If we set the uncertainty from impurities at twice the fractional level of impurities, the largest error (0.12\%) comes from the Arajs and Dunmyre data on iron. Given the purity of the Fe and Ni samples used, we assign no additional systematic error beyond that already determined from the spread in the data. We will revisit the effects of impurities once again in the determination of the spin component of the magnetization.

Another source of impurities generally not accounted for in assays is the surface oxidation. Iron oxides such as Fe$_3$O$_4$, have a much smaller magnetization than pure Fe. Alex Gray's group at Temple University  took XMCD measurements for us at the Advanced Light Source on a pure Fe foil which we provided from our M\o ller target materials. These measurements, which probe the material surface to a depth of a few nanometers, showed clear evidence of surface oxidation in spite of their highly specular appearance. This suggests that foils nearing micron level thickness could have surface contamination from oxides at the 0.1\% level. We expect that using clean foils with no surface oxidation apparent to the naked eye and with a thickness of 10~$\mu$m will render this source of uncertainty negligible at the $\ll$0.1\% level. 

\subsubsection{Nuclear contribution to the magnetic moment}
Discussion of the nuclear contribution to the magnetic moment appears to be absent from the literature on magnetization measurements. This is most likely due to the suppression of the nuclear magneton relative to the Bohr magneton by the electron to proton mass ratio ($\mu_B/\mu_N=m_p/m_e$), a factor of about 1/2000. However, in the determination of target polarization for the M\o ller polarimeter, effects at the 0.1\% level require consideration. In the nucleus spins are paired in such a way that all even-even nuclei have zero spin. Fortunately, the isotopic distribution of iron (26 protons) is such that 97.9\% of natural iron is from even-even isotopes. The single even-odd naturally occuring isotope $^{57}$Fe has a negligible nuclear spin of 0.09$\mu_N$\cite{Locher1965}. For nickel (28 protons) the situation is also favorable with natural nickel being composed of 98.9\% even-even isotopes. This gives us another two orders of magnitude suppression and renders the nuclear spin contribution completely negligible. However, for cobalt (27 protons), the only stable isotope has a nuclear spin of 4.63~$\mu_N$, potentially creating errors at the 0.2\% level and adding another reason not to use Co foil.   

\subsubsection{Defects from target irradiation}
Another potential source of systematic error in determining target saturation magnetization is the effect of radiation damage. If a sufficient fraction of lattice sites are dislodged/damaged this could potentially change the target saturation polarization. We estimated the radiation damage by integrating the Mott scattering cross section from momentum transfer of infinity down to the threshold set by the permanent lattice displacement energy (nuclear recoil energy of 40~eV) weighting the cross section by the number of additional atoms that are dislodged by the initial atom using the NRT method to estimate the displacements per atom \cite{Norgett1975}. This produced a total cross section of order 100 barns. While this effective displacement cross section is relatively large, it would take more than 100 years in our typical 1~$\mu$A beam for a significant fraction of the target lattice sites to be displaced. This is consistent with non-observation (to the best of our knowledge) of such an effect in any M\o ller polarimeter worldwide. Given that we have not observed such an effect directly at Jefferson Lab in our extensive use of precision M\o ller polarimeters in both Halls A and C, and that our order of magnitude  estimate suggests insignificant fractional damage, we have chosen not to add an additional systematic error to account for radiation damage.

\subsection{Determination of $g^{\prime}$ and the spin component of magnetization}
Magnetization arises from a combination of spin and orbital contributions. In ferromagnetic materials, the orbital component is suppressed  or ``quenched" compared to the spin. To find the spin polarization of the target foils we must determine the spin fraction of the magnetization. The spin component of the magnetization can be determined from measurements of $g^{\prime}$, the total g-factor for atomic electrons which can be obtained from magnetomechanical experiments utilizing the Einstein-de Haas effect or the Barnett effect.\footnote{The Einstein-de Haas effect (rotation by magnetization) is the rotation of a macroscopic body in a magnetic field when the field is reversed\cite{Richardson1908, Scott1962}. The Barnett effect (magnetization by rotation) is the converse, the production of a magnetic field by rotation of a macroscopic body\cite{Barnett1909, Barnett1944}.} In general, the $g$-factor is related the to gyromagnetic ratio $\gamma$ of a charged body as 
\begin{equation}
\gamma=g\frac{\mu_B}{\hbar},
\label{eq:gyro}
\end{equation}
where $\mu_B$ is the Bohr magneton.\footnote{In early publications sometimes the gyromagnetic ratio is given as $\rho=L/M$ the ratio of the angular momentum to the magnetic moment where at other times it is defined in the usual way as the reciprocal $\gamma=1/\rho=M/L$.} The electron has two $g$-factors which we refer to as $g_{S}\approx2$ for its spin, and $g_{L}=1$ for its orbital motion. For atoms having both orbital and spin angular momentum, $g^{\prime}$ is a linear combination of $g_{S}$ and $g_{L}$, which is not known {\it a priori} and must be determined from measurement.

In publications from the early to middle 1900s, $g_{S}$ was assumed to be exactly 2 where we now know it to be (up to a sign) the most precisely measured scientific constant $g_{S}=2.00231930436256(35)$. In most cases, this $0.1\%$ difference is not consequential, but for the level of precision we are trying to reach, this is not negligible and care must be taken to track down wherever 2 has been substituted for $g_{S}$. 

The relationship of $g^{\prime}$ to the magnetic moment contribution is often given in the literature following the example of Kittel\cite{Kittel1949} in the following form: \cite{Meyer1961, Smit1959}
\begin{equation}
g^{\prime}=\frac{2(M_{S}+M_{L})}{M_{S}+2M_{L}}=\frac{2M_{\rm tot}}{M_{\rm tot}+M_{L}},
\label{eq:gprime_approx}
\end{equation}
where $M_{\rm tot}$ is the total magnetization. $M_{L}$ and $M_{S}$  are the components of magnetization arising from orbital and spin magnetic moments respectively. This expression immediately leads to the expression of orbital and spin contributions to the magnetic moment as \cite{deBever1997}
\begin{equation}
\frac{M_{L}}{M_{\rm tot}}=\frac{2-g^{\prime}}{g^{\prime}}, ~~\frac{M_{S}}{M_{\rm tot}}=1-\frac{M_{L}}{M_{\rm tot}}.
\label{eq:frac_gprime_approx}
\end{equation}

The gyromagnetic ratio, $\gamma$ is defined as the ratio of the magnetic moment of a particle or body to its angular momentum. In measurements of $g^\prime$ where magnetization and angular momentum of macroscopic bodies are directly measured, the gyromagnetic ratio is determined as 
\[
\gamma=\frac{M}{J},
\]
where $M$ and $J$ are the projections of $\textbf{M}$ and $\textbf{J}$ along the direction of magnetization. We can divide these into their spin and orbital components:
\[
M=M_{L}+M_{S},~~~~J=J_{L}+J_{S},
\]
where the subscripts $L$ and $S$ refer to orbital and spin respectively.  At the atomic level the magnetic moment $\textbf{M}$ is related to the orbital and spin angular momentum as $\textbf{M}_{S}=g_{S}\mu_B\textbf{S}/\hbar$ and $\textbf{M}_{L}=g_{L}\mu_B\textbf{L}/\hbar$, such that a unit of spin angular momentum yields $g_S/g_L$ more magnetic moment than a unit of orbital angular momentum. This holds also at the macroscopic level so that we can write
\begin{equation}
\gamma=g^{\prime}\frac{\mu_B}{\hbar},~~~~~g^\prime=\frac{M_{\rm tot}}{M_{S}/g_{S}+M_{L}/g_{L}}.
\end{equation}
To high precision $g_{L}=1$ yielding \footnote{There is a small correction to $g_{L}$ that arises from the finite mass of the nucleus at the order of the ratio of the electron mass to that of the nucleus ($\sim$1$\times10^{-5}$)\cite{Phillips1949}. This is two orders of magnitude below the correction considered here of $(g_S-2)/g_S$ and will be neglected.}
\begin{equation}
g^{\prime}=\frac{M_{\rm tot}}{M_{S}/g_{S}+M_{L}}=\frac{g_SM_{\rm tot}}{M_{S}+g_{S}M_{L},}.
\label{eq:gprime_exact}
\end{equation}
from which we recover Eq. \ref{eq:gprime_approx} if we substitute $g_{S}=2$. Eq. \ref{eq:gprime_exact} is the exact form which should be used in this analysis. Furthermore, the exact form of Eq. \ref{eq:frac_gprime_approx} is the slightly more complicated
\begin{equation}
\frac{M_{L}}{M_{\rm tot}}=\frac{g_{S}-g^{\prime}}{g^{\prime}(g_{S}-1)}.
\label{eq:frac_gprime_exact}
\end{equation}
This gives for the spin component
\begin{equation}
\frac{M_{S}}{M_{\rm tot}}=1-\frac{M_{L}}{M_{\rm tot}}=\frac{g_{S}(g^{\prime}-1)}{g^{\prime}(g_{S}-1)},
\label{eq:frac_sp_exact}
\end{equation}
which decreases the spin contribution to the total magnetization compared to Eq. \ref{eq:frac_gprime_approx} by $~0.11\%$.

\subsubsection{$g^{\prime}$ for Fe}
The most precise measurments of $g^{\prime}$ come from measurements of the gyromagnetic ratio of iron using the Einstein-de Haas effect. These magnetomechanical experiments are highly elaborate requiring high precision to observe the tiny effects of interest. The Einstein-de Haas experiments are simple in principle: a sample is suspended from a torsion pendulum along the axis of a magnetic field. Upon reversal of the field a small torque on the sample is measured primarily due to reversal of the valence electron spins. In practice, these experiments are highly technical since the torques on the sample from the Earth's magnetic field can be 7-8 orders of magnitude larger than the torques from spin reversal\cite{Scott1962}. Elaborate coil setups were utilized to cancel the Earth's field along with any stray magnetic fields in the region and isolation systems incorporated to keep the sample free from interference from outside vibrations. The gyromagnetic ratio was then determined from the measured ratio of the angular momentum to the magnetic moment. Similarly complex systems were used in the experiments which measured the Barnett effect. In these experiments a relatively large sample was rotated and the change in magnetic flux measured in a system of pickup coils. 

\begin{figure}[t]
\centering
\includegraphics[width=0.8\textwidth]{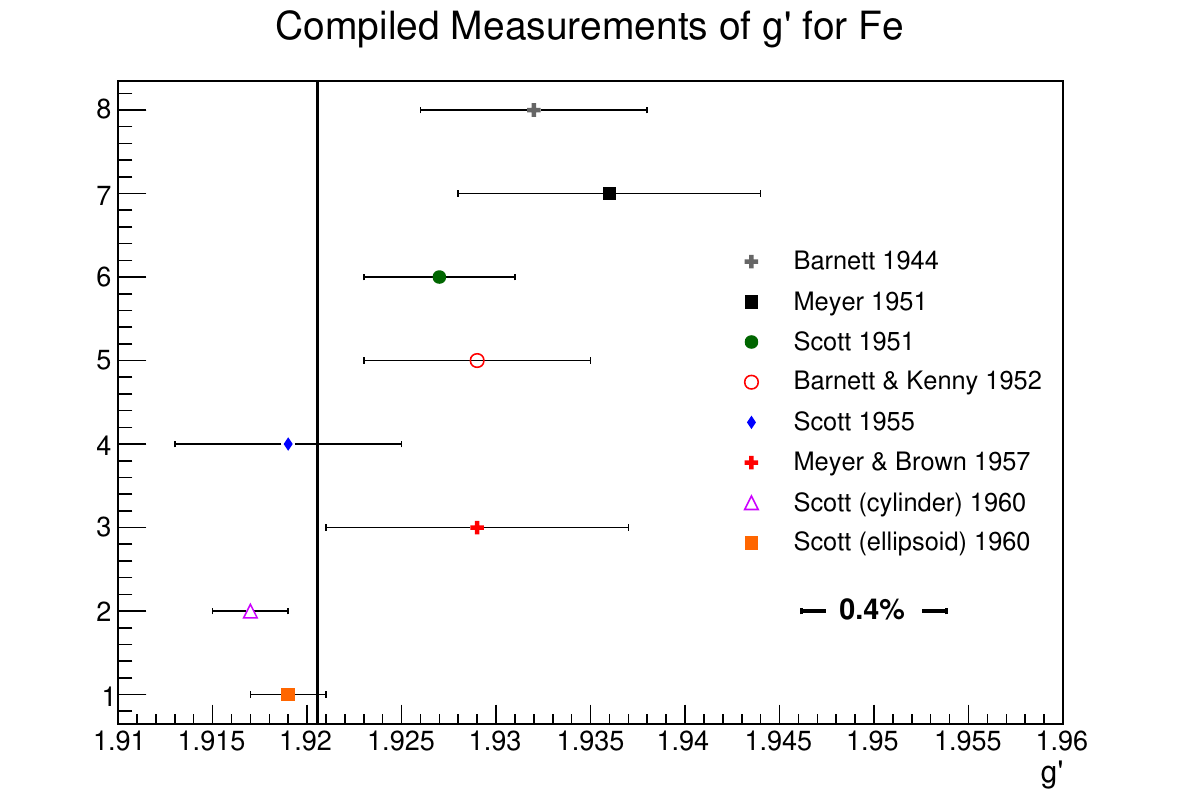}
\caption{Values of $g^{\prime}$ for iron as determined by various experiments between 1940 and 1960. The naive constant fit to these data is given by the vertical black line whose value is $g^{\prime}=1.9206$.}
\label{fig:gprime_world_data_Fe}
\end{figure}
A compilation of $g^{\prime}$ measurements on iron from magnetomechanical experiments is shown in Fig. \ref{fig:gprime_world_data_Fe}. These data were taken from compilations in two papers\footnote{There are two inconsistencies between these references\cite{Scott1962,Meyer1961}. 1. Table 1 of \cite{Meyer1961} has Barnett 1941 $\rho e/mc=1.035$ ($g^\prime=1.932$) which comes from averaging measurements using the Einstein-de Haas and Barnett effects. Scott seems to only use Barnett's measurements of the Einstein-de Haas effect and quotes Barnett's measurement as $g^\prime=1.938$. We retain Barnett's average of the two methods. 2. Scott \cite{Scott1962} gives Meyer's 1957 value for Fe as $g^\prime$=1.932, whereas Meyer \cite{Meyer1961} uses 1.929. We use Meyer's value.} 
 by G. Scott in 1962\cite{Scott1962} and Meyer and Asch in 1961\cite{Meyer1961}. For reference, the data included in these compilations comes from \cite{Barnett1944,Scott1951,Barnett1952,Meyer1957,Scott1960}. The final two measurements done by G. Scott are by far the most precise. It is clear given the fit probability of 0.004 and from discussions of how the uncertainties were determined, that the error bars do not in all cases reflect the actual systematic error, which, in at least some of the measurements, is underestimated. The most accurate measurements were made by Scott, who without stated justification, concludes that his most recent measurement of $g^{\prime}=1.919\pm0.002$ on a prolate ellipsoid sample is the best value to use for iron \cite{Scott1960, Scott1962} even though he measured  $g^{\prime}=1.917\pm0.002$ on a cylindrical sample using the same apparatus. It is likely that he regarded the ellipsoid-shaped sample more accurate because of the uniformity of the internal magnetic field this shape produces. It is worth noting that his latest value  $g^{\prime}=1.919$ appears to be the value taken as standard in the literature (see for example \cite{Wohlfarth1980,Bonnenberg1986}). It not clear what systematics may be at play here (sample purity, shape, porosity, preparation/annealing process). 
 
For the three samples used in the measurements $g^{\prime}$ of Fe, the sample purities were as follows: 
\begin{itemize}
\item{Scott cylinder 99.94\% with primary impurities O(0.04\%), C(0.005\%), N(0.004\%), S(0.003\%) and Ni(0.0015\%) \cite{Scott1951}}
\item{Scott ellipsoid, 99.89\% with primary impurities Ni(0.05\%), Si(0.01\%), O(0.005\%), Co(0.005\%) \cite{Scott1960}}
\item{ Meyer 1957, 99.9\% with primary impurities Mn(0.042\%), S(0.029\%), Si(0.02\%) \cite{Meyer1957}} 
\end{itemize}

Scott carefully measured the effect of mixing the ferromagnetic elements Fe, Co and Ni and since their $g^{\prime}$ values are all within 5\% of each other trace amounts of impurities ($<$1\%) from of Ni and Co in Fe will have negligible effect on the value of  $g^{\prime}$ (see Fig 1 of \cite{Scott1969}). There is little guidance in the literature for the effect of trace amounts of O, Mn, N, C and S on $g^{\prime}$ for Fe making it difficult to set the scale for such errors. However, Ladislav Pust {\it et al.} found very little difference in the related quantity spectroscopic $g$ between pure Fe and that with 3\% Si by weight\cite{Pust1984}. We will see in the coming paragraphs that the spectroscopic g-factor is inversely related to $g^{\prime}$ such that if one increases, the other decreases and vice versa. 

An error-weighted fit to these data gives a result of 1.9206$\pm$0.0012. However, the $\chi^2$/NDF is 2.41 indicating that systematic errors have been underestimated. Following the example of the Particle Data Group (see Sec. 5.2.2 of \cite{PDG2018}), and inflating each of the error bars by $\sqrt{\chi^2/{\rm NDF}}=1.553$ to give a $\chi^2/$NDF of unity (p-value = 0.43) yields an error of 0.0019 or $\pm$0.10\%.

Related to $g^{\prime}$ is the spectroscopic $g$-factor often referred to as $g$ from ferromagnetic resonance (FMR) experiments\footnote{For a simple explanation of FMR see http://www.physik.fu-berlin.de/einrichtungen/ag/ag-kuch/research/techniques/fmr/index.html}. FMR works by placing a ferromagnetic sample in a resonant microwave cavity. The cavity is placed in a uniform magnetic field at right angles to the direction of propagation of the microwaves. A microwave source feeds the cavity and a detector monitors the energy coming out of the cavity. When the magnetic field is turned on, the magnetic moments of the atoms will begin to precess around the direction of the applied magnetic field with a frequency that depends on the effective magnetic field $H_{\rm eff}$ and the $g$-factor of the sample material as follows:
\begin{equation}
\hbar\omega=g\mu_BH_{\rm eff}
\label{eq:fmr}
\end{equation}
where $H_{\rm eff}$, the effective magnetic field depends on the applied magnetic field strength as well as the magnetization, shape and relative alignment of the specimen (see \cite{Kittel1949, Smit1959} for a more detailed explanation). The magnetic field strength is then swept over a range until the resonance condition is met where the precession frequency matches that of the microwave cavity. At resonance a drop in power exiting the cavity will be observed due to the energy being absorbed by the sample. Spectroscopic $g$ is determined by measuring the magnetic field which excites this resonance. For a time it was thought that spectroscopic $g$ and $g^{\prime}$ were the same i.e. that spectroscopic and magnetomechanical experiments were measuring the same $g$-factor until Kittel (1949)\cite{Kittel1949} and Van Vleck (1950)\cite{Vleck1950} independently showed that these are related but not identical quantities. In the case of spectroscopic $g$, the lattice momentum offsets the intrinsic orbital momentum so that the total angular momentum is approximately equal to the spin contribution\cite{Kittel1949, Reck1969}. Therefore, spectroscopic $g$ is given by
\begin{equation}
g\left(\frac{\mu_B}{\hbar}\right)=\frac{M_{L}+M_{S}}{S},
\end{equation}
where $S$ is the electron spin. To a good approximation it can be shown that $g=\frac{2M_{\rm tot}}{M_{\rm tot}-{M_{L}}}$ where $g^{\prime}$ is given approximately by Eq. \ref{eq:gprime_approx}. Thus, the orbital component increases the magnitude of $g$ and decreases $g^{\prime}$. Using these equations we can easily derive what is known as the Kittel-Van Vleck relationship 
\begin{equation}
\frac{1}{g}+\frac{1}{g^{\prime}}=1.
\label{eq:kittelvanvleck}
\end{equation}

\begin{figure}[h]
\centering
\includegraphics[width=0.8\textwidth]{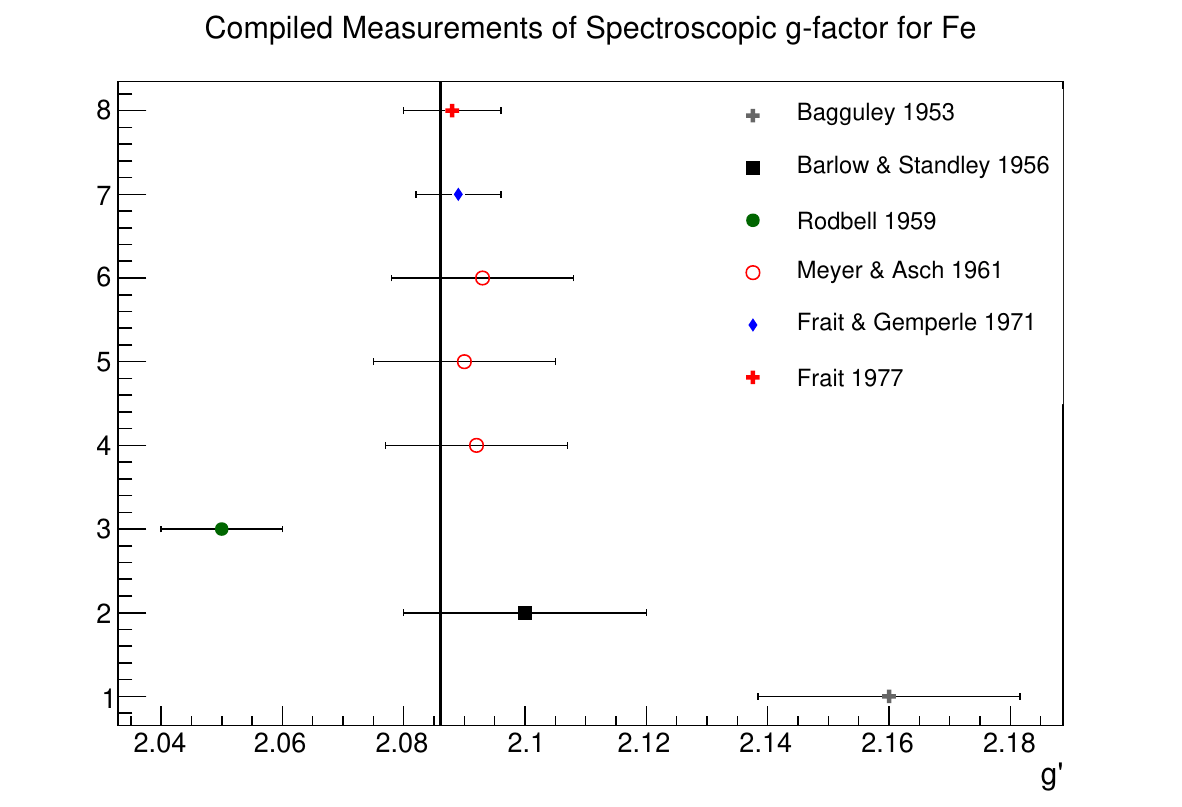}
\caption{Values of spectroscopic $g$ as determined by various experiments over two decades. The error-weighted fit to these data is given by the vertical black line whose value is $g=2.086$.}
\label{fig:gfactor_world_data_Fe}
\end{figure}

Although this relationship is approximate and should not be considered valid below the $\pm$0.1\% level, it has been shown to work quite well in the literature (see for example Fig. 1 of \cite{Meyer1961}). Therefore, we can utilize spectroscopic measurements of $g$ to further check our value of $g^{\prime}$. Figure \ref{fig:gfactor_world_data_Fe} shows a compilation of measurements of $g$ for iron. A simple error-weighted fit to these data gives a value of $g=2.086\pm0.004$. Using Eq. \ref{eq:kittelvanvleck} gives $g^{\prime}=1.921$ in precise agreement with the error weight fit to $g^{\prime}$ from magnetomechanical experiments. While we cannot place the same confidence in this derived value of  $g^{\prime}$ as the direct measurements, it is reassuring that determinations from completely different techniques appear to be consistent.

{\bf Recommendation for Fe:} In light of these findings we recommend using the value of the simple error-weighted fit with an inflated systematic error to reflect the tension in the world data: $\bm{ g^{\prime}=1.9206\pm0.0019}$. The 0.0019 error comes from inflating the error reported by the fit by 55.3\% which is required to remove the tension in the data and give a $\chi^2/$NDF of 1. The systematic error from impurities is assumed to be included in this uncertainty. This choice places Scott's recommended value of $g^{\prime}=1.919\pm0.002$ measured on an ellipsoid Fe sample \cite{Scott1962} comfortably within $1\sigma$ but his earlier measurement on a cylindrical sample $1.9\sigma$ off. 

\subsubsection{ $ g^{\prime}$ for Ni}
A number of measurements of $g^{\prime}$ for nickel were performed by A. J. Meyer {\it et al.}, G. G. Scott {\it et al.} and S. Barnett {\it et al.} during the 1950's. At first there were striking differences in the values found for nickel ranging from 1.83 to $>$1.99. Furthermore, the measurements of spectrocopic $g$ from resonance experiments gave a much lower value of $g^{\prime}$ using the Eq. \ref{eq:kittelvanvleck}. A couple of systematic errors in the measurement techniques of both Meyer and Scott were pointed out by Brown which brought the data into much better agreement\cite{Scott1962}. However, a considerable inconsistency remained between the measurement of Barnett {\it et al.} and that of Scott and Meyer. Barnett determined $g^{\prime}\approx1.91$ compared to the 4\% lower $g^{\prime}\approx1.84$ found by Meyer and Scott\cite{Meyer1961,Scott1962}. To investigate the possible reasons for this discrepancy, Meyer measured the Curie temperature and the saturation magnetization of the Ni samples used in each of the measurements. Whereas Scott and Meyer had used nearly pure Ni, Barnett's sample had 1.4\% impurities. The presence of these impurities significantly changed the magnetic properties of his Ni sample such that the Curie temperature was reduced from 360$^{\circ}$C for pure Ni to 285$^{\circ}$C and the saturation magnetization increased from 58.90 to 71.04 (in units of abamp cm$^3$/g)\cite{Scott1962}. Scott concludes that this stark shift in magnetic properties makes Barnett's measurements ``difficult to retain''\cite{Scott1962}. However, this discrepancy provides evidence that the presence of certain impurities can have a significant effect on the measurement of $g^\prime$.

Scott performed a series of four measurements on the same Ni sample in 1952, 1953, 1955 and 1960 and concluded that $g^{\prime}=1.835\pm0.002$\cite{Scott1962}. Meyer {\it et al.} also measured $g^{\prime}$ for different Ni samples in 1957 and 1958 finding 1.852$\pm$0.009 and 1.845$\pm$0.007\cite{Meyer1961}. An error-weighted fit to these values gives $g^{\prime}=1.8365\pm0.0019$ with a $\chi^2$/NDF of 2.5. 

The impurities in the samples used are as follows:
\begin{itemize}
\item{Scott: 99.82\% Ni with main impurities Si(0.1\%), Fe(0.032\%), Mn(0.030\%), and C(0.01\%)\cite{ScottSep1955}}
\item{Meyer, 1957: 99.9\% Ni with impurities not provided\cite{Meyer1957}}
\item{Meyer, 1958: 99.99\% with negligible impurities\cite{Meyer1961}}
\end{itemize}

Looking at the impurities in Scott's sample, we can rule out the effects of Fe and Mn as contributing significantly to a systematic offset using the data in \cite{Standley1955, Scott1969}. With carbon impurities at 0.01\% this can be considered negligible. Meyer's analysis of the magnetic properties of the Ni sample used by Scott showed that although the saturation magnetization was changed insignificantly, the Curie temperature decreased by 11$^{\circ}$C. Since we were not able to locate data to calibrate the effect of Si impurities at 0.1\% in Ni, a similar approach to that used for the Fe data will be used here. Inflating the error bars on each of the three data points by 1.581 gives a best fit of $g^{\prime}=1.8365\pm$0.0030 with a p-value of 0.37.  


\begin{figure}[h]
\centering
\includegraphics[width=0.8\textwidth]{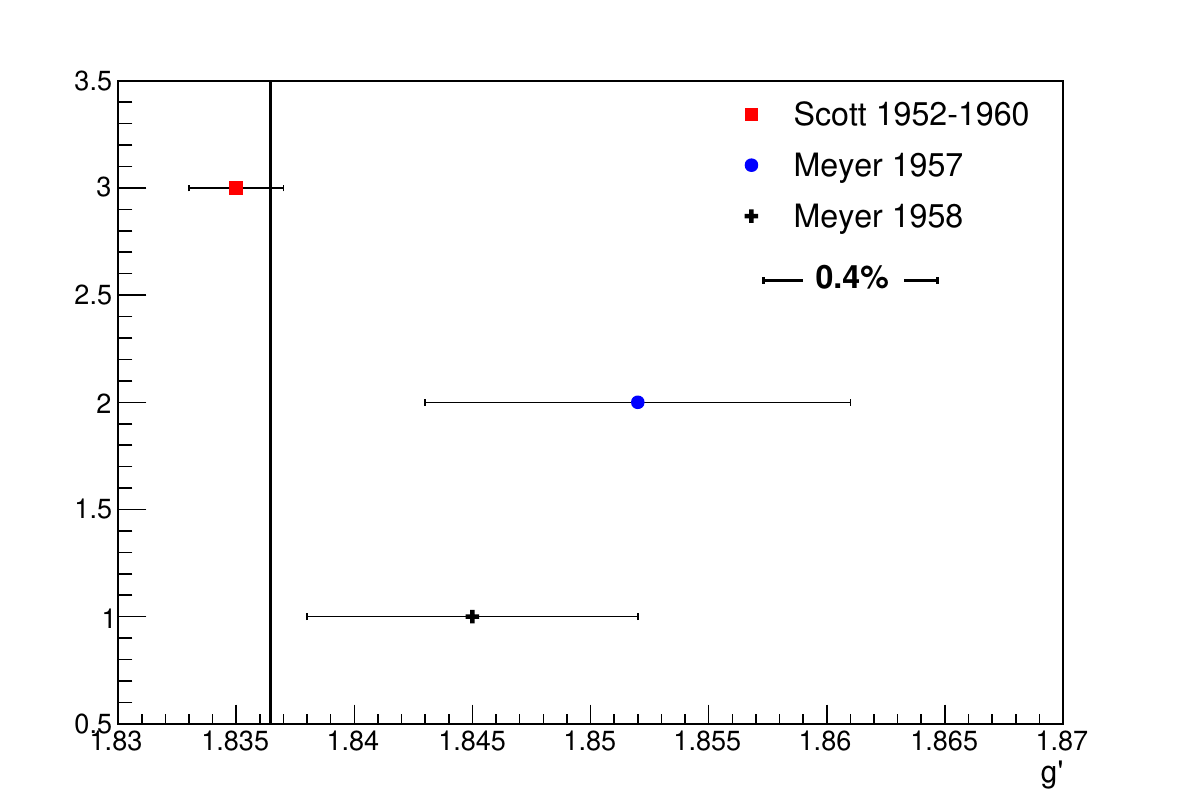}
\caption{Values of $g^{\prime}$ for nickel as determined by various experiments between 1950 and 1960. The systematic error on Scott's value as proposed in the text is shown. The error-weighted fit to these data using the proposed error given by the vertical line is $g^{\prime}=1.8365\pm0.0036$.}
\label{fig:gprime_world_data_Ni}
\end{figure}

Once again we can use measurements of the spectroscopic $g$-factor from magnetic resonance experiments and Eq. \ref{eq:kittelvanvleck} as an independent check of our proposed value of $g^{\prime}$. Table II. of Meyer and Asch \cite{Meyer1961} provided a compilation of $g$-factors measured in magnetic resonance experiments and concluded that for nickel $g=2.185\pm0.010$ which translates into $g^{\prime}=1.844\pm0.008$ in good agreement with our proposed value.

{\bf Recommendation for Ni:} in light of these findings we recommend using the value {$\bm{ g^{\prime}=1.8365\pm0.0030}$ for nickel. The value comes from an error-weighted fit to Scott's and Meyer's measured values after increasing each of the error bars by 1.581 to accommodate for the underestimated systematic uncertainty. 

\subsubsection{\label{sec:gp_temp_dep}Temperature dependence of $g^{\prime}$}
The measurements of $g^{\prime}$ used in this analysis have all been at room temperature which is not well-defined but is broadly accepted to be near 20$^{\circ}$C give or take a few degrees. Although the target foils in the M\o ller polarimeter will generally be at room temperature, during measurements with a typical 1~$\mu$A of beam on target, the foils will heat up by 10-15 degrees Celsius as we saw in section \ref{sec:target_heating}. This raises the question of whether or not the room temperature values of $g^{\prime}$ are sufficiently accurate during measurements at elevated temperatures.

The temperature dependence of saturation magnetization arising from spin waves was discussed in section \ref{sec:sat_mag_THdep}. If this change in saturation magnetization results in a change of the fraction of magnetization arising from orbital and spin components, this would necessarily imply a change in $g^\prime$. Conversely, a temperature-independent $g^\prime$ would imply that spin waves proportionately decrease both the orbital and spin components of magnetization. 

In Kittel's 1949 paper on the relation of $g$ and $g^{\prime}$, he discusses the temperature dependence of $g^{\prime}$ and suggests there is not enough data to make conclusions\cite{Kittel1949}. Since then several measurements have been made of $g$ across a broad temperature range for the ferromagnetic elements and alloys. These experiments, which measure $g$ since it is a technically much easier measurement than $g^{\prime}$, particularly with changing temperatures, are typically at the 1-2\% precision level. However, a change in $g$ indicates the inverse change in the  $g^{\prime}$ by Eq. \ref{eq:kittelvanvleck}. A nice summary of these measurements is found in \cite{borovik1988}.

It is worth noting that in all cases where pure Ni and Fe were measured, the $g$-factor was always found to be constant within experimental errors, typically at the 1-2\% level. However, for alloys, this is not always the case with variations of several percent being observed (see for example \cite{Gadsden1978, Shanina1998}).

In two cases, extremely accurate measurements were made across a broad temperature range, one for pure Ni and the other for 97\% Fe. The first of these was by G. Dewar {\it et al.} in 1977 on pure nickel foil of 20~$\mu$m thickness. They found $g=2.187\pm0.005$ constant over the temperature range 20-364$^{\circ}$C\cite{Dewar1977}. This constitutes a 0.23\% test of temperature dependence over a range much larger than we care about. The second experiment in 1981 by Ladislav Pust and Zdenek Frait measured the $g$-factor of Fe-3wt\%Si in the temperature range from 3.5 to 300 K to be constant at $g=2.0793\pm0.0005$\cite{Pust1981}. The extreme accuracy of their measurement allowed them to probe the temperature dependence of $g$ at the 0.02\% level and they conclude that there is no evidence of temperature dependence across the temperature range they measured. The plot from their paper showing the measurement of $g$ with temperature is shown in Fig. \ref{fig:Frait1981}. A summary of the various measurements of $g$ is provided in Table \ref{tab:gfactor_Tdep}.

Thus, there is strong evidence that spectroscopic $g$ and by extension $g^{\prime}$ are, in fact, highly constant for nickel and iron well below their Curie temperatures. This implies that the spin-wave correction does not significantly alter the fraction of magnetic moment arising from orbital and spin contributions for these two ferromagnetic elements. We will revisit spin waves in the context of the field-dependence of $g^\prime$, but we conclude that it is safe to proceed with confidence using the room temperature measurements of $g^{\prime}$ with negligible error.

\begin{table}[h]
\caption{\label{tab:gfactor_Tdep}Results of experiments measuring the spectroscopic $g$-factor as a function of temperature for various ferromagnetic materials. Without exception all consider the $g$-factor to be constant within error.}
\begin{center}
\begin{tabular}{|r|l|l|l|l|}\hline
Publication&Year&Material&$g$-factor&Temp. ($^{\circ}$C)\\\hline
Frait {\it et al.} \cite{Pust1981}&1981&Fe-3wt\%Si&2.0793$\pm$0.0005&$-$270 to 27\\
Haraldson {\it et al.} \cite{Haraldson1981}&1981&Ni&2.20$\pm$0.02&20 to 358\\
Gadsden {\it et al.} \cite{Gadsden1978}&1978&Ni&2.20&$-$269 to 20\\
Dewar {\it et al.} \cite{Dewar1977}&1977&Ni&2.187$\pm$0.005&20 to 364\\
Bastian {\it et al.} \cite{Bastian1976_2}&1976&Ni-Fe alloys&const. $\pm$1\% &20 to $>$300\\
Rodbell \cite{Rodbell1964}&1964&Ni&2.22$\pm$0.03&$-$140 to 360\\
Rodbell \cite{Rodbell1959}&1959&Fe&2.05$\pm$0.01&$-$196 to 850\\
Standley {\it et al.}\cite{Standley1955}&1955&Ni&2.17-2.18&20 to 200\\
Bagguley {\it et al.}\cite{Bagguley1954}&1954&Ni&2.22$\pm$0.02&20 to 600\\
Bloembergen \cite{Bloembergen1950}&1950&Ni&2.20$\pm$1-2\%&24 to 358\\\hline

\end{tabular}
\end{center}
\end{table}

\begin{figure}[h]
\centering
\includegraphics[width=0.8\textwidth]{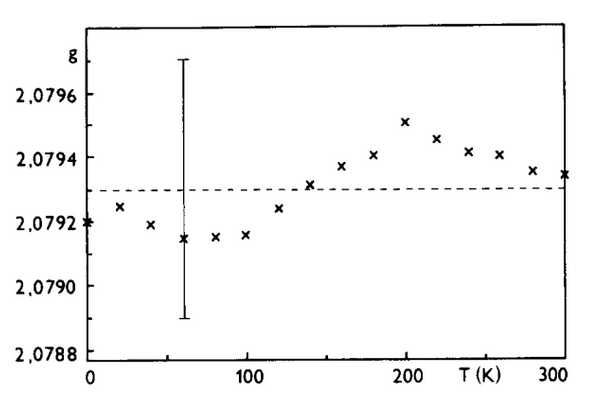}
\caption{Plot of $g$-values vs. temperature taken from \cite{Pust1981}. The vertical bar denotes the accuracy of these values ($\pm0.0004$). }
\label{fig:Frait1981}
\end{figure}

\subsubsection{Magnetic field dependence of $g^{\prime}$}
In the 1950's while Scott was performing precise measurements of $g^{\prime}$, he initially found that $g^{\prime}$ decreased at very low fields and asymptotically approached a larger constant value at higher fields. He published three papers documenting the low-field behavior of $g^{\prime}$ for nickel and iron and alloys of the two \cite{ScottAug1955, ScottSep1955, ScottAug1956}. In 1960, he found that this low-field behavior was due to a systematic error in his measurement technique\cite{Scott1960}. After improving the technique and re-measuring, he concluded that, in fact, $g^\prime$ is independent of applied field for Ni and Fe over the range of fields he was measuring. His setup utilized a solenoid with a total area 78000~cm$^2$ which he energized with 1-16~mA producing fields as high as 40 gauss. Although these fields were sufficient to induce significant magnetization in the elongated samples, the high currents only induced magnetizations approaching half the level of saturation magnetization. Here we look at evidence to demonstrate that $g^\prime$ remains field-independent in the several tesla applied field region where the M\o ller polarimeter operates.

FMR measurements of spectroscopic $g$ are taken with the sample at saturation magnetization where the magnetization is well-determined from the literature and the $g$-factor can be calculated (see Eq. \ref{eq:fmr}). The frequency independence of the $g$-factor often tested in the literature is simultaneously a test of the magnetic field-dependence of $g$ since the frequency is a function of the effective field, $H_{eff}$.

In 1971, Z. Frait and R. Gemperle measured the $g$-factor of single iron crystals across a range of frequencies from 12 to 70~GHz requiring a broad range of static magnetic fields\cite{Frait1971} which roughly corresponds to applied fields from  0.08~T to 1.6~T (for details on converting between resonance frequency and applied field see Kittel\cite{Kittel1948}). They found that $g=2.089\pm0.007$ and that it  is frequency independent over this range within their experimental error ($\pm$0.33\%). In 1977, Z. Frait published an FMR measurement of $g=2.088\pm0.008$ for pure polycrystalline iron at three frequencies, 26~GHz (at 0.32~T), 36~GHz (at 0.57~T) and 70~GHz (at 1.53~T)\cite{Frait1977}. Once again he concluded that within experimental error this value is frequency independent, constituting a high-field test of field dependence on $g$ for iron. Unfortunately, Pust {\it et al.} make no mention of frequency-dependence in their $\pm$0.024\% measurement of the $g$-factor of Fe-3wt\%Si even though their results were averages of four different frequencies, 36~GHz, 70~GHz, 86~GHz and 95~GHz\cite{Pust1981}. 

For nickel the data are less precise but point to the same conclusion that $g$ is field-independent. In 1950 Bloembergen measured the $g$-factor of nickel to be 2.23 at 9.05~GHz with a field of 0.116~T and 2.24 at 22.44~GHz with a magnetic field of 0.54~T. These values are equal within the error of the experiment. In 1959, Rodbell found that for nickel $g$ was constant at the 0.5\% level over a range of magnetic fields up to 0.3~T\cite{Rodbell1959}. In 1965, Frait found that $g$ was independent of frequency for pure nickel at the 2\% level over a range of frequencies from 8.5~GHz to 72~GHz (roughly corresponding to applied fields of 0.1 T - 2.4 T). He also found that an alloy consisting of 42\% Fe and 58\% Ni was independent of frequency over the same range at the 1\% level\cite{Frait1965}. Finally, as we saw earlier in section \ref{sec:gp_temp_dep} the value of $g^{\prime}$ for nickel derived from high-field measurements of $g$ agrees well within error with the direct measurements at low field, providing further evidence of the validity of the asymptotic value of $g^{\prime}$ for nickel. 

Although we found no field-dependence of $g^\prime$ for Fe and Ni in the literature, the evidence is not sufficiently precise to rule out 0.1\% level changes at high field. Given this consideration we chose to place an upper limit on the field dependence using measurements of high-field susceptibility as we outline next.

Given that $g^\prime$ provides a measure of the fraction of the magnetization from orbital and spin contributions (see Eq. \ref{eq:frac_gprime_exact}) any field dependence of $g$ or $g^\prime$ is a signal that the fractional contribution from spin is field-dependent. In section \ref{sec:gp_temp_dep}, we concluded that the spin-wave correction did not significantly alter $g^\prime$ as evidenced from the temperature independence of $g$; however, there are other field-dependent contributions to magnetization which can be separated from the spin-wave contribution by either going to the high-field or low temperature regime where spin-wave contributions are negligible. The linear increase of magnetization with applied field in the high-field region is referred to as the high-field susceptibility $\chi_{\rm HF}(H)=\partial M/\partial H$. $\chi_{\rm HF}$ is composed of both orbital and spin contributions\cite{Herring1966, Stoelinga1966, Foner1969}. Some attempts have been made to calculate the relative contributions of the orbital and spin to the high-field susceptibility\cite{Yasui1971}. An upper limit on the field dependence of the spin fraction can be made by assigning the full high-field change in magnetization solely to a spin or to an orbital contribution.  Tables \ref{tab:Fe_chiHF} and \ref{tab:Ni_chiHF} list 5 measurements of the high-field susceptibility for Fe and Ni respectively. The average of the five measurements is 0.0065~emu/(g kOe) for Fe and 0.0025~emu/(g kOe) for Ni. The error is given by the product of $\chi_{\rm HF}$ and the internal field in the foil divided by the saturation magnetization. For Fe (Ni) foils the field is set to 4~(2)~T giving an internal field of 18.4~(13.8)~kOe. With saturation magnetization for Fe~(Ni) of 218~(55.2)~emu/g this gives a final percent error of 0.055 (0.063)\%. We add this additional error in quadrature with the error in the orbital fraction propagated from the uncertainty in $g^\prime$.
\begin{table}[h]
\caption{\label{tab:Fe_chiHF}Measurements of $\chi_{\rm HF}$ in the high-field and/or low temperature regime for iron. The measurement by Herring {\it et al.} is almost 3 times larger than the average of the others. The reason for this is not clear, but this measurement was conservatively retained in the average. The ``Error" column is the percent contribution to the magnetization at an applied field of 4~T.}
\begin{center}
\begin{tabular}{lccc}\hline
Publication&Material&$\chi_{\rm HF} \left(\frac{\rm emu}{ \rm g~kOe}\right)$&Error \%\\\hline
Herring {\it et al.} 1966 \cite{Herring1966}&Fe+4\%Si&0.0140&0.118\\
Foner {\it et al.} 1966 \cite{Foner1966}&Fe&0.0051&0.043\\
Stoelinga {\it et al.} 1966 \cite{Stoelinga1966}&Fe&0.0041&0.035\\
Foner {\it et al.} 1969 \cite{Foner1969}&Fe&0.0055&0.046\\
Pauthenet {\it et al.} 1982 \cite{PauthenetMar1982}&Fe&0.0036&0.031\\\hline
~&Average&0.0065&0.055\\
\end{tabular}
\end{center}
\end{table}

\begin{table}[h]
\caption{\label{tab:Ni_chiHF}Measurements of $\chi_{\rm HF}$ in the high-field and/or low temperature regime for nickel. Once again, the measurement by Herring {\it et al.} is 3 times larger than the average of the others. The ``Error" column is the percent contribution to the magnetization at an applied field of 2~T.}
\begin{center}
\begin{tabular}{lccc}\hline
Publication&Material&$\chi_{\rm HF} \left(\frac{\rm emu}{ \rm g~kOe}\right)$&Error \%\\\hline
Herring {\it et al.} 1966 \cite{Herring1966}&Ni&0.0056&0.141\\
Foner {\it et al.} 1966 \cite{Foner1966}&Ni&0.0012&0.031\\
Stoelinga {\it et al.} 1966 \cite{Stoelinga1966}&Ni&0.0023&0.057\\
Foner {\it et al.} 1969 \cite{Foner1969}&Ni&0.0019&0.048\\
Pauthenet {\it et al.} 1982 \cite{PauthenetMar1982}&Ni&0.0016&0.040\\\hline
~&Average&0.0025&0.063\\
\end{tabular}
\end{center}
\end{table}

\FloatBarrier
\section{Calculation of Target Polarization}\label{sec:final_calc}
We are now in a position to calculate the final target polarization and the uncertainty on the value. Tables \ref{tab:final_errors_Fe} and \ref{tab:final_errors_Ni} provide the data for Fe and Ni respectively. The values for magnetization and polarization are calculated for applied magnetic fields of 4~T and 2~T for Fe and Ni foils respectively. In the calculation of target polarization by deBever {\it et al.} \cite{deBever1997}, the magnetic moment of an electron is assumed to be 1~$\mu_B$, which is an approximation valid in the limit that $g_{S}=2$ since $\mu_e=\frac{g_S}{2}\mu_B$. Thus the magnetic moment of an electron is approximately 1.00116$\mu_B$ and this approximation introduces an error at the 0.1\% level. 

Temperature corrections due to target heating are calculated for a 1~$\mu$A beam load. To first order, increasing the beam load linearly increases the temperature correction whereas increasing target thickness leaves the temperature unchanged. This insensitivity of temperature to thickness is due to the assumption of a good thermal contact with an infinite heat sink at the foil edge. Under these assumptions, the increased conduction of the thicker foil offsets the additional heat load. Therefore, increasing foil thickness is the better choice for increasing scattering rates. 
\begin{table}[h]
\caption{\label{tab:final_errors_Fe}Summary of values and errors involved in calculating the target polarization for Fe foils.}
\begin{center}
\begin{tabular}{|l|l|l|c|}\hline
Quantity&T=294 K&T=307 K&Unit\\\hline
Saturation magnetization $M_s$ &218.04(44)&217.73(45)&emu/g\\
Saturation magnetization $M_s$&2.1803(44)&2.1771(45)&$\mu_B$/atom\\
$g^{\prime}$&1.9206(19)&1.9206(19)&$-$\\
Orbital fraction: $\frac{M_{L}}{M_{\rm tot}}=\frac{g_{S}-g^\prime}{g^\prime(g_{S}-1)}$&0.0425(10)&0.0425(10)&$-$\\
Spin component: $M_S\left(1-\frac{M_{L}}{M_{\rm tot}}\right)$&2.0877(47)&2.0847(48)&$\mu_B$/atom\\
Average electron magnetization&0.08030(18)&0.08018(19)&$\mu_B$\\
Average electron polarization&0.08020(18)&0.08009(19)&$-$\\\hline
\end{tabular}
\end{center}
\end{table}

\begin{table}[h]
\caption{\label{tab:final_errors_Ni}Summary of values and errors involved in calculating the target polarization for Ni foils.}
\begin{center}
\begin{tabular}{|l|l|l|c|}\hline
Quantity&T=294 K&T=307 K&Unit\\\hline
Saturation magnetization $M_s$ &55.24(11)&54.91(15)&emu/g\\
Saturation magnetization $M_s$&0.5806(12)&0.5771(16)&$\mu_B$/atom\\
$g^{\prime}$&1.8365(30)&1.8365(30)&$-$\\
Orbital fraction: $\frac{M_{L}}{M_{\rm tot}}=\frac{g_{S}-g^\prime}{g^\prime(g_{S}-1)}$&0.0901(18)&0.0901(18)&$-$\\
Spin component: $M_S\left(1-\frac{M_{L}}{M_{\rm tot}}\right)$&0.5283(15)&0.5251(18)&$\mu_B$/atom\\
Average electron magnetization&0.018867(53)&0.018753(63)&$\mu_B$\\
Average electron polarization&0.018845(53)&0.018731(63)&$-$\\\hline
\end{tabular}\end{center}
\end{table}

Thus we have demonstrated that the saturation polarization of an Fe target can be determined to $\pm$0.23\% under a 1~$\mu$A beam load, typical for Hall A at Jefferson Lab. For the same conditions the polarization for a Ni target can be determined to $\pm$0.33\% . However, it is important to verify that the target truly is saturated at the magnetic field settings for a given experiment. Further discussion of this topic including sensitivity to target alignment and flatness are a topic for an additional publication. 

A total of $\pm$0.25\% is currently alotted in our proposed uncertainty budget for target polarization for the MOLLER experiment, implying that we must demonstrate that we are within 0.1\% of saturation for an iron target.  Although Ni polarization uncertainty is significantly higher than Fe, a significant contribution that can be greatly reduced comes from the heating correction. The heating correction for Ni is much larger than for Fe due to its low Curie temperature. Reducing the current from 1 to 0.3~$\mu$A for a Ni foil reduces the overall systematic error from $\pm$0.33\% to $\pm$0.28\%. Thus, a single precision, low current measurement on a Ni foil could be of value for crosschecking the systematic error on the polarization for Fe.

\section{Concluding Discussion}\label{conclusions}
The polarization of a saturated ferromagnetic target has been calculated for both nickel and iron foils. With the stringent demands of the proposed MOLLER experiment, it seemed wise to revisit the study of Fe target polarization by deBever {\it et al.}\cite{deBever1997}. A different approach was taken than that in \cite{deBever1997} where instead of using the saturation magnetization value at 0~K and then correcting back to room temperature, measured values of magnetization were taken at or near room temperature. A small error was found in the magnetic field correction in equation (3) of \cite{deBever1997} where the applied magnetic field was used instead of the internal magnetic field, introducing a small error of about 0.1\%. Using the approximation $g_S=2$ also introduced further errors of order 0.1\% in \cite{deBever1997}. 

Using measurements of magnetization and $g^\prime$ we calculate the saturation target polarization for Fe  foils at room temperature with 4~T fields applied normal to the foil to be 0.08020$\pm$0.00018. For Ni foils under a 2~T applied field, the saturation polarization is 0.018845$\pm$0.000053. We are optimistic that utilizing an Fe foil target will allow us to reach our uncertainty goal of $\pm$0.25\% for target polarization including all uncertainties.

Recent evidence from measurement in Hall A revealed our sensitivity to wrinkles in the foil and raised questions about how well our foils were aligned normal to the holding field. Deviations of the foil surface from normality make it more difficult to reach saturation which is the only place where polarization is known with high accuracy. Further studies will be needed and are ongoing to determine the level of foil flatness required and our sensitivity to foil alignment angle. These are topics of discussion for a future publication.

We would like to thank Silviu Covrig of Jefferson Lab for cross-checking our simple target heating model with his ANSYS-Fluent software package. We also acknowledge the support of the U.S. Department of Energy. This material is based upon the work supported by the U.S. Department of Energy, Office of Science, Office of Nuclear Physics Contract
No. DE-AC05-06OR23177. Temple University also acknowledges the support of the U.S. Department of Energy, Office of Science, Office of Nuclear Physics under contract  DE-SC0020422.
\FloatBarrier



\bibliographystyle{elsarticle-num} 


\bibliography{bibliography}

\begin{thebibliography}{10}
\expandafter\ifx\csname url\endcsname\relax
  \def\url#1{\texttt{#1}}\fi
\expandafter\ifx\csname urlprefix\endcsname\relax\def\urlprefix{URL }\fi
\expandafter\ifx\csname href\endcsname\relax
  \def\href#1#2{#2} \def\path#1{#1}\fi

\bibitem{MOLLER2014}
{The MOLLER Collaboration}, The {MOLLER} experiment: An ultra-precise
  measurement of the weak mixing angle using {M\o}ller scattering (2014).
\newblock \href {http://arxiv.org/abs/1411.4088} {\path{arXiv:1411.4088}}.

\bibitem{SoLID2019}
{The SoLID collaboration},
  \href{https://hallaweb.jlab.org/12GeV/SoLID/files/solid-precdr-Nov2019.pdf}{{SoLID}
  ({S}olenoidal {L}arge {I}ntensity {D}evice) updated preliminary conceptual
  design report}.
\newline\urlprefix\url{https://hallaweb.jlab.org/12GeV/SoLID/files/solid-precdr-Nov2019.pdf}

\bibitem{deBever1997}
L.~de~Bever, J.~Jourdan, M.~Loppacher, S.~Robinson, I.~Sick, J.~Zhao,
  \href{http://www.sciencedirect.com/science/article/pii/S0168900297009613}{A
  target for precise {M}\o ller polarimetry}, Nuclear Instruments and Methods
  in Physics Research Section A: Accelerators, Spectrometers, Detectors and
  Associated Equipment 400~(2) (1997) 379 -- 386.
\newblock \href
  {https://doi.org/http://dx.doi.org/10.1016/S0168-9002(97)00961-3}
  {\path{doi:http://dx.doi.org/10.1016/S0168-9002(97)00961-3}}.
\newline\urlprefix\url{http://www.sciencedirect.com/science/article/pii/S0168900297009613}

\bibitem{Swartz1995}
M.~Swartz, H.~Band, F.~Decker, P.~Emma, M.~Fero, R.~Frey, R.~King, A.~Lath,
  T.~Limberg, R.~Prepost, P.~Rowson, B.~Schumm, M.~Woods, M.~Zolotorev,
  \href{https://www.sciencedirect.com/science/article/pii/0168900295003843}{Observation
  of target electron momentum effects in single-arm møller polarimetry},
  Nuclear Instruments and Methods in Physics Research Section A: Accelerators,
  Spectrometers, Detectors and Associated Equipment 363~(3) (1995) 526--537.
\newblock \href {https://doi.org/https://doi.org/10.1016/0168-9002(95)00384-3}
  {\path{doi:https://doi.org/10.1016/0168-9002(95)00384-3}}.
\newline\urlprefix\url{https://www.sciencedirect.com/science/article/pii/0168900295003843}

\bibitem{PREX2021}
D.~Adhikari, {\it et. al.},
  \href{https://link.aps.org/doi/10.1103/PhysRevLett.126.172502}{Accurate
  determination of the neutron skin thickness of $^{208}\mathrm{Pb}$ through
  parity-violation in electron scattering}, Phys. Rev. Lett. 126 (2021) 172502.
\newblock \href {https://doi.org/10.1103/PhysRevLett.126.172502}
  {\path{doi:10.1103/PhysRevLett.126.172502}}.
\newline\urlprefix\url{https://link.aps.org/doi/10.1103/PhysRevLett.126.172502}

\bibitem{Kraftmakher2005}
Y.~Kraftmakher, \href{https://doi.org/10.1119/1.1994857}{Spontaneous
  magnetization of ferromagnets}, American Journal of Physics 73~(12) (2005)
  1191--1194.
\newblock \href {http://arxiv.org/abs/https://doi.org/10.1119/1.1994857}
  {\path{arXiv:https://doi.org/10.1119/1.1994857}}, \href
  {https://doi.org/10.1119/1.1994857} {\path{doi:10.1119/1.1994857}}.
\newline\urlprefix\url{https://doi.org/10.1119/1.1994857}

\bibitem{KittelOct1949}
C.~Kittel, \href{https://link.aps.org/doi/10.1103/RevModPhys.21.541}{Physical
  theory of ferromagnetic domains}, Rev. Mod. Phys. 21 (1949) 541--583.
\newblock \href {https://doi.org/10.1103/RevModPhys.21.541}
  {\path{doi:10.1103/RevModPhys.21.541}}.
\newline\urlprefix\url{https://link.aps.org/doi/10.1103/RevModPhys.21.541}

\bibitem{Bloch1930}
F.~Bloch, \href{http://dx.doi.org/10.1007/BF01339661}{Zur theorie des
  ferromagnetismus}, Zeitschrift f{\"u}r Physik 61~(3) (1930) 206--219.
\newblock \href {https://doi.org/10.1007/BF01339661}
  {\path{doi:10.1007/BF01339661}}.
\newline\urlprefix\url{http://dx.doi.org/10.1007/BF01339661}

\bibitem{Kittel1951}
C.~Herring, C.~Kittel,
  \href{https://link.aps.org/doi/10.1103/PhysRev.81.869}{On the theory of spin
  waves in ferromagnetic media}, Phys. Rev. 81 (1951) 869--880.
\newblock \href {https://doi.org/10.1103/PhysRev.81.869}
  {\path{doi:10.1103/PhysRev.81.869}}.
\newline\urlprefix\url{https://link.aps.org/doi/10.1103/PhysRev.81.869}

\bibitem{Dyson1956_2}
F.~J. Dyson,
  \href{https://link.aps.org/doi/10.1103/PhysRev.102.1230}{Thermodynamic
  behavior of an ideal ferromagnet}, Phys. Rev. 102 (1956) 1230--1244.
\newblock \href {https://doi.org/10.1103/PhysRev.102.1230}
  {\path{doi:10.1103/PhysRev.102.1230}}.
\newline\urlprefix\url{https://link.aps.org/doi/10.1103/PhysRev.102.1230}

\bibitem{Foner1969}
S.~Foner, A.~J. Freeman, N.~A. Blum, R.~B. Frankel, E.~J. McNiff, H.~C.
  Praddaude, \href{https://link.aps.org/doi/10.1103/PhysRev.181.863}{High-field
  studies of band ferromagnetism in {Fe and Ni} by {M}\"ossbauer and magnetic
  moment measurements}, Phys. Rev. 181 (1969) 863--882.
\newblock \href {https://doi.org/10.1103/PhysRev.181.863}
  {\path{doi:10.1103/PhysRev.181.863}}.
\newline\urlprefix\url{https://link.aps.org/doi/10.1103/PhysRev.181.863}

\bibitem{PauthenetMar1982}
R.~Pauthenet, \href{http://dx.doi.org/10.1063/1.330694}{Spin‐waves in nickel,
  iron, and yttrium‐iron garnet}, Journal of Applied Physics 53~(3) (1982)
  2029--2031.
\newblock \href {https://doi.org/10.1063/1.330694}
  {\path{doi:10.1063/1.330694}}.
\newline\urlprefix\url{http://dx.doi.org/10.1063/1.330694}

\bibitem{Dyson1956}
F.~J. Dyson, \href{https://link.aps.org/doi/10.1103/PhysRev.102.1217}{General
  theory of spin-wave interactions}, Phys. Rev. 102 (1956) 1217--1230.
\newblock \href {https://doi.org/10.1103/PhysRev.102.1217}
  {\path{doi:10.1103/PhysRev.102.1217}}.
\newline\urlprefix\url{https://link.aps.org/doi/10.1103/PhysRev.102.1217}

\bibitem{Keffer1966}
F.~Keffer, Spin Waves, Springer, Berlin, Heidelberg, 1966, pp. 1--273.
\newblock \href {https://doi.org/https://doi.org/10.1007/978-3-642-46035-7_1}
  {\path{doi:https://doi.org/10.1007/978-3-642-46035-7_1}}.

\bibitem{PauthenetNov1982}
R.~Pauthenet, \href{http://dx.doi.org/10.1063/1.330287}{Experimental
  verification of spin‐wave theory in high fields (invited)}, Journal of
  Applied Physics 53~(11) (1982) 8187--8192.
\newblock \href {https://doi.org/10.1063/1.330287}
  {\path{doi:10.1063/1.330287}}.
\newline\urlprefix\url{http://dx.doi.org/10.1063/1.330287}

\bibitem{Graham1982}
C.~D.~G. Jr., \href{http://dx.doi.org/10.1063/1.330695}{Iron and nickel as
  magnetization standards}, Journal of Applied Physics 53~(3) (1982)
  2032--2034.
\newblock \href {https://doi.org/10.1063/1.330695}
  {\path{doi:10.1063/1.330695}}.
\newline\urlprefix\url{http://dx.doi.org/10.1063/1.330695}

\bibitem{Skomski2007}
R.~Skomski, G.~C. Hadjipanayis, D.~J. Sellmyer, Effective demagnetizing factors
  of complicated particle mixtures, IEEE Transactions on Magnetics 43~(6)
  (2007) 2956--2958.
\newblock \href {https://doi.org/10.1109/TMAG.2007.893798}
  {\path{doi:10.1109/TMAG.2007.893798}}.

\bibitem{Owen1954}
E.~A. Owen, D.~M. Jones,
  \href{http://stacks.iop.org/0370-1301/67/i=6/a=302}{Effect of grain size on
  the crystal structure of cobalt}, Proceedings of the Physical Society.
  Section B 67~(6) (1954) 456.
\newline\urlprefix\url{http://stacks.iop.org/0370-1301/67/i=6/a=302}

\bibitem{Case1966}
W.~E. Case, R.~D. Harrington,
  \href{http://nvlpubs.nist.gov/nistpubs/jres/70C/jresv70Cn4p255_A1b.pdf}{Calibration
  of vibrating-sample magnetometers}, Journal of Research of the National
  Bureau of Standards-C Engineering and Instrumentation 70C~(4) (1966)
  255--262.
\newline\urlprefix\url{http://nvlpubs.nist.gov/nistpubs/jres/70C/jresv70Cn4p255_A1b.pdf}

\bibitem{Crangle1971}
J.~Crangle, G.~M. Goodman,
  \href{http://rspa.royalsocietypublishing.org/content/321/1547/477}{The
  magnetization of pure iron and nickel}, Proceedings of the Royal Society of
  London A: Mathematical, Physical and Engineering Sciences 321~(1547) (1971)
  477--491.
\newblock \href {https://doi.org/10.1098/rspa.1971.0044}
  {\path{doi:10.1098/rspa.1971.0044}}.
\newline\urlprefix\url{http://rspa.royalsocietypublishing.org/content/321/1547/477}

\bibitem{Danan1968}
H.~Danan, A.~Herr, A.~J.~P. Meyer,
  \href{http://dx.doi.org/10.1063/1.2163571}{New determinations of the
  saturation magnetization of nickel and iron}, Journal of Applied Physics
  39~(2) (1968) 669--670.
\newblock \href {https://doi.org/10.1063/1.2163571}
  {\path{doi:10.1063/1.2163571}}.
\newline\urlprefix\url{http://dx.doi.org/10.1063/1.2163571}

\bibitem{Aldred1975}
A.~T. Aldred,
  \href{https://link.aps.org/doi/10.1103/PhysRevB.11.2597}{Temperature
  dependence of the magnetization of nickel}, Phys. Rev. B 11 (1975)
  2597--2601.
\newblock \href {https://doi.org/10.1103/PhysRevB.11.2597}
  {\path{doi:10.1103/PhysRevB.11.2597}}.
\newline\urlprefix\url{https://link.aps.org/doi/10.1103/PhysRevB.11.2597}

\bibitem{Weiss1929}
P.~Weiss, R.~Forrer, \href{https://doi.org/10.1051/anphys/192910120279}{La
  saturation absolue des ferromagnétiques et les lois d'approche en fonction
  du champ et de la température}, Ann. Phys. 10~(12) (1929) 279--372.
\newblock \href {https://doi.org/10.1051/anphys/192910120279}
  {\path{doi:10.1051/anphys/192910120279}}.
\newline\urlprefix\url{https://doi.org/10.1051/anphys/192910120279}

\bibitem{Sanford1941}
R.~L. Sanford, E.~G. Bennett,
  \href{http://nistdigitalarchives.contentdm.oclc.org/cdm/ref/collection/p13011coll6/id/104774}{A
  determination of the magnetic saturation induction of iron at room
  temperature}, NIST Journal of Research (Jan 1941).
\newline\urlprefix\url{http://nistdigitalarchives.contentdm.oclc.org/cdm/ref/collection/p13011coll6/id/104774}

\bibitem{Danan1959}
H.~Danan, \href{https://hal.archives-ouvertes.fr/jpa-00236018}{On the
  interpretation of the magnetization measurements of pure polycrystalline iron
  and nickel in the vicinity of saturation}, {J. Phys. Radium} 20~(2-3) (1959)
  203--207.
\newblock \href {https://doi.org/10.1051 / jphysrad: 01959002002-3020300}
  {\path{doi:10.1051 / jphysrad: 01959002002-3020300}}.
\newline\urlprefix\url{https://hal.archives-ouvertes.fr/jpa-00236018}

\bibitem{Arajs1967}
S.~Arajs, G.~R. Dunmyre, \href{http://dx.doi.org/10.1002/pssb.19670210117}{A
  note on the consistency of values of the spontaneous or saturation
  magnetization of polycrystalline iron and nickel at 298 {$^\circ$K}}, Physica
  Status Solidi (b) 21~(1) (1967) 191--195.
\newblock \href {https://doi.org/10.1002/pssb.19670210117}
  {\path{doi:10.1002/pssb.19670210117}}.
\newline\urlprefix\url{http://dx.doi.org/10.1002/pssb.19670210117}

\bibitem{Behrendt1972}
D.~R. Behrendt, D.~E. Hegland,
  \href{https://ntrs.nasa.gov/search.jsp?R=19720015089}{Saturation
  magnetization of polycrystalline iron}, Tech. rep., NASA (Apr 1972).
\newline\urlprefix\url{https://ntrs.nasa.gov/search.jsp?R=19720015089}

\bibitem{Stoner1950}
C.~S. Edmund, \href{https://doi.org/10.1088/0034-4885/13/1/304}{Ferromagnetism:
  magnetization curves}, Reports on Progress in Physics 13~(1) (1950) 83--183.
\newblock \href {https://doi.org/10.1088/0034-4885/13/1/304}
  {\path{doi:10.1088/0034-4885/13/1/304}}.
\newline\urlprefix\url{https://doi.org/10.1088/0034-4885/13/1/304}

\bibitem{Foner1956}
S.~Foner, \href{https://link.aps.org/doi/10.1103/PhysRev.101.1648}{Hall effect
  and magnetic properties of armco iron}, Phys. Rev. 101 (1956) 1648--1652.
\newblock \href {https://doi.org/10.1103/PhysRev.101.1648}
  {\path{doi:10.1103/PhysRev.101.1648}}.
\newline\urlprefix\url{https://link.aps.org/doi/10.1103/PhysRev.101.1648}

\bibitem{Cullity2008}
B.~D. Cullity, C.~D. Graham, Introduction to Magnetic Materials, 2nd Edition,
  Wiley-IEEE Press, 2008.

\bibitem{Myers1951}
H.~P. Myers, W.~Sucksmith,
  \href{http://rspa.royalsocietypublishing.org/content/207/1091/427}{The
  spontaneous magnetization of cobalt}, Proceedings of the Royal Society of
  London A: Mathematical, Physical and Engineering Sciences 207~(1091) (1951)
  427--446.
\newblock \href
  {http://arxiv.org/abs/http://rspa.royalsocietypublishing.org/content/207/1091/427.full.pdf}
  {\path{arXiv:http://rspa.royalsocietypublishing.org/content/207/1091/427.full.pdf}},
  \href {https://doi.org/10.1098/rspa.1951.0132}
  {\path{doi:10.1098/rspa.1951.0132}}.
\newline\urlprefix\url{http://rspa.royalsocietypublishing.org/content/207/1091/427}

\bibitem{JonesDC2022}
J.~D. C.,
  \href{https://moller.jlab.org/DocDB/0008/000874/001/TargetHeating.pdf}{Solutions
  to the heat equation for circular target foil heating by an electron beam for
  uniform and gaussian beam distributions}.
\newline\urlprefix\url{https://moller.jlab.org/DocDB/0008/000874/001/TargetHeating.pdf}

\bibitem{Arajs1964}
S.~Arajs, R.~V. Colvin,
  \href{http://dx.doi.org/10.1063/1.1702873}{Ferromagnetic‐paramagnetic
  transition in iron}, Journal of Applied Physics 35~(8) (1964) 2424--2426.
\newblock \href {https://doi.org/10.1063/1.1702873}
  {\path{doi:10.1063/1.1702873}}.
\newline\urlprefix\url{http://dx.doi.org/10.1063/1.1702873}

\bibitem{Arajs1963}
S.~Arajs, R.~Colvin,
  \href{http://www.sciencedirect.com/science/article/pii/0022369763902427}{Paramagnetism
  of polycrystalline nickel}, Journal of Physics and Chemistry of Solids
  24~(10) (1963) 1233 -- 1237.
\newblock \href
  {https://doi.org/http://dx.doi.org/10.1016/0022-3697(63)90242-7}
  {\path{doi:http://dx.doi.org/10.1016/0022-3697(63)90242-7}}.
\newline\urlprefix\url{http://www.sciencedirect.com/science/article/pii/0022369763902427}

\bibitem{Arajs1965}
S.~Arajs, \href{http://dx.doi.org/10.1063/1.1714136}{Paramagnetic behavior of
  nickel just above the ferromagnetic curie temperature}, Journal of Applied
  Physics 36~(3) (1965) 1136--1137.
\newblock \href {https://doi.org/10.1063/1.1714136}
  {\path{doi:10.1063/1.1714136}}.
\newline\urlprefix\url{http://dx.doi.org/10.1063/1.1714136}

\bibitem{Luborsky1980}
F.~E. Luborsky, J.~L. Walter, E.~P. Wohlfarth,
  \href{http://stacks.iop.org/0305-4608/10/i=5/a=024}{The saturation
  magnetisation, curie temperature and size effect of amorphous iron alloys},
  Journal of Physics F: Metal Physics 10~(5) (1980) 959.
\newline\urlprefix\url{http://stacks.iop.org/0305-4608/10/i=5/a=024}

\bibitem{Ahern1958}
S.~A. Ahern, M.~J.~C. Martin, W.~Sucksmith,
  \href{http://www.jstor.org/stable/100593}{The spontaneous magnetization of
  nickel+copper alloys}, Proceedings of the Royal Society of London. Series A,
  Mathematical and Physical Sciences 248~(1253) (1958) 145--152.
\newline\urlprefix\url{http://www.jstor.org/stable/100593}

\bibitem{Locher1965}
P.~R. Locher, S.~Geschwind,
  \href{https://link.aps.org/doi/10.1103/PhysRev.139.A991}{Electron-nuclear
  double resonance of {${\mathrm{Fe}}^{57}$} in {MgO}}, Phys. Rev. 139 (1965)
  A991--A994.
\newblock \href {https://doi.org/10.1103/PhysRev.139.A991}
  {\path{doi:10.1103/PhysRev.139.A991}}.
\newline\urlprefix\url{https://link.aps.org/doi/10.1103/PhysRev.139.A991}

\bibitem{Norgett1975}
M.~Norgett, M.~Robinson, I.~Torrens,
  \href{https://www.sciencedirect.com/science/article/pii/0029549375900357}{A
  proposed method of calculating displacement dose rates}, Nuclear Engineering
  and Design 33~(1) (1975) 50--54.
\newblock \href {https://doi.org/https://doi.org/10.1016/0029-5493(75)90035-7}
  {\path{doi:https://doi.org/10.1016/0029-5493(75)90035-7}}.
\newline\urlprefix\url{https://www.sciencedirect.com/science/article/pii/0029549375900357}

\bibitem{Richardson1908}
O.~W. Richardson,
  \href{https://link.aps.org/doi/10.1103/PhysRevSeriesI.26.248}{A mechanical
  effect accompanying magnetization}, Phys. Rev. (Series I) 26 (1908) 248--253.
\newblock \href {https://doi.org/10.1103/PhysRevSeriesI.26.248}
  {\path{doi:10.1103/PhysRevSeriesI.26.248}}.
\newline\urlprefix\url{https://link.aps.org/doi/10.1103/PhysRevSeriesI.26.248}

\bibitem{Scott1962}
G.~G. {Scott}, {Review of Gyromagnetic Ratio Experiments}, Reviews of Modern
  Physics 34 (1962) 102--109.
\newblock \href {https://doi.org/10.1103/RevModPhys.34.102}
  {\path{doi:10.1103/RevModPhys.34.102}}.

\bibitem{Barnett1909}
S.~J. Barnett, \href{https://doi.org/10.1126/science.30.769.413}{On
  magnetization by angular acceleration} (Sep. 1909).
\newblock \href {https://doi.org/10.1126/science.30.769.413}
  {\path{doi:10.1126/science.30.769.413}}.
\newline\urlprefix\url{https://doi.org/10.1126/science.30.769.413}

\bibitem{Barnett1944}
S.~J. Barnett, \href{http://www.jstor.org/stable/20023462}{New researches on
  magnetization by rotation and the gyromagnetic ratios of ferromagnetic
  substances}, Proceedings of the American Academy of Arts and Sciences 75~(5)
  (1944) 109--129.
\newline\urlprefix\url{http://www.jstor.org/stable/20023462}

\bibitem{Kittel1949}
C.~Kittel, \href{https://link.aps.org/doi/10.1103/PhysRev.76.743}{On the
  gyromagnetic ratio and spectroscopic splitting factor of ferromagnetic
  substances}, Phys. Rev. 76 (1949) 743--748.
\newblock \href {https://doi.org/10.1103/PhysRev.76.743}
  {\path{doi:10.1103/PhysRev.76.743}}.
\newline\urlprefix\url{https://link.aps.org/doi/10.1103/PhysRev.76.743}

\bibitem{Meyer1961}
A.~J.~P. Meyer, G.~Asch,
  \href{http://dx.doi.org/10.1063/1.2000457}{Experimental $g^\prime$ and $g$
  values of {Fe, Co, Ni}, and their alloys}, Journal of Applied Physics 32~(3)
  (1961) S330--S333.
\newblock \href {https://doi.org/10.1063/1.2000457}
  {\path{doi:10.1063/1.2000457}}.
\newline\urlprefix\url{http://dx.doi.org/10.1063/1.2000457}

\bibitem{Smit1959}
J.~Smit, H.~Wijn, Ferrites, Eindhoven: Philips Technical Library, 1959.

\bibitem{Phillips1949}
M.~Phillips, \href{https://link.aps.org/doi/10.1103/PhysRev.76.1803}{The effect
  of nuclear motion on atomic magnetic moments}, Phys. Rev. 76 (1949)
  1803--1804.
\newblock \href {https://doi.org/10.1103/PhysRev.76.1803}
  {\path{doi:10.1103/PhysRev.76.1803}}.
\newline\urlprefix\url{https://link.aps.org/doi/10.1103/PhysRev.76.1803}

\bibitem{Scott1951}
G.~G. Scott, \href{https://link.aps.org/doi/10.1103/PhysRev.82.542}{A precise
  mechanical measurement of the gyromagnetic ratio of iron}, Phys. Rev. 82
  (1951) 542--547.
\newblock \href {https://doi.org/10.1103/PhysRev.82.542}
  {\path{doi:10.1103/PhysRev.82.542}}.
\newline\urlprefix\url{https://link.aps.org/doi/10.1103/PhysRev.82.542}

\bibitem{Barnett1952}
S.~J. Barnett, G.~S. Kenny,
  \href{https://link.aps.org/doi/10.1103/PhysRev.87.723}{Gyromagnetic ratios of
  iron, cobalt, and many binary alloys of iron, cobalt, and nickel}, Phys. Rev.
  87 (1952) 723--734.
\newblock \href {https://doi.org/10.1103/PhysRev.87.723}
  {\path{doi:10.1103/PhysRev.87.723}}.
\newline\urlprefix\url{https://link.aps.org/doi/10.1103/PhysRev.87.723}

\bibitem{Meyer1957}
{Meyer, André J.P.}, {Brown, Sheldon},
  \href{https://doi.org/10.1051/jphysrad:01957001803016100}{Nouvelles mesures
  des rapports gyromagnétiques du fer et du nickel}, J. Phys. Radium 18~(3)
  (1957) 161--168.
\newblock \href {https://doi.org/10.1051/jphysrad:01957001803016100}
  {\path{doi:10.1051/jphysrad:01957001803016100}}.
\newline\urlprefix\url{https://doi.org/10.1051/jphysrad:01957001803016100}

\bibitem{Scott1960}
G.~G. Scott,
  \href{https://link.aps.org/doi/10.1103/PhysRev.119.84}{Gyromagnetic ratios of
  {Fe and Ni}}, Phys. Rev. 119 (1960) 84--85.
\newblock \href {https://doi.org/10.1103/PhysRev.119.84}
  {\path{doi:10.1103/PhysRev.119.84}}.
\newline\urlprefix\url{https://link.aps.org/doi/10.1103/PhysRev.119.84}

\bibitem{Wohlfarth1980}
E.~Wohlfarth,
  \href{http://www.sciencedirect.com/science/article/pii/S1574930405801166}{Chapter
  1 iron, cobalt and nickel}, Handbook of Ferromagnetic Materials 1 (1980) 35.
\newblock \href
  {https://doi.org/http://dx.doi.org/10.1016/S1574-9304(05)80116-6}
  {\path{doi:http://dx.doi.org/10.1016/S1574-9304(05)80116-6}}.
\newline\urlprefix\url{http://www.sciencedirect.com/science/article/pii/S1574930405801166}

\bibitem{Bonnenberg1986}
D.~Bonnenberg, K.~A. Hempel, H.~Wijn,
  \href{https://doi.org/10.1007/10311893_25}{1.2.1.2.4 Atomic magnetic moment,
  magnetic moment density, $g$ and $g^\prime$ factor}, Springer Berlin
  Heidelberg, Berlin, Heidelberg, 1986, pp. 174--188.
\newblock \href {https://doi.org/10.1007/10311893_25}
  {\path{doi:10.1007/10311893_25}}.
\newline\urlprefix\url{https://doi.org/10.1007/10311893_25}

\bibitem{Scott1969}
G.~G. Scott, H.~W. Sturner,
  \href{https://link.aps.org/doi/10.1103/PhysRev.184.490}{Magnetomechanical
  ratios for {Fe-Co} alloys}, Phys. Rev. 184 (1969) 490--491.
\newblock \href {https://doi.org/10.1103/PhysRev.184.490}
  {\path{doi:10.1103/PhysRev.184.490}}.
\newline\urlprefix\url{https://link.aps.org/doi/10.1103/PhysRev.184.490}

\bibitem{Pust1984}
L.~Půst, Z.~Frait,
  \href{http://dx.doi.org/10.1002/pssb.2221220218}{Low-temperature {FMR and
  FMAR} measurements of tetal single crystals. {I. General Consideration,
  Experimental Techniques}}, Physica Status Solidi (b) 122~(2) (1984) 535--541.
\newblock \href {https://doi.org/10.1002/pssb.2221220218}
  {\path{doi:10.1002/pssb.2221220218}}.
\newline\urlprefix\url{http://dx.doi.org/10.1002/pssb.2221220218}

\bibitem{PDG2018}
M.~{\it et al.}. Tanabashi,
  \href{https://link.aps.org/doi/10.1103/PhysRevD.98.030001}{Review of particle
  physics}, Phys. Rev. D 98 (2018) 030001.
\newblock \href {https://doi.org/10.1103/PhysRevD.98.030001}
  {\path{doi:10.1103/PhysRevD.98.030001}}.
\newline\urlprefix\url{https://link.aps.org/doi/10.1103/PhysRevD.98.030001}

\bibitem{Vleck1950}
J.~H. Van~Vleck,
  \href{https://link.aps.org/doi/10.1103/PhysRev.78.266}{Concerning the theory
  of ferromagnetic resonance absorption}, Phys. Rev. 78 (1950) 266--274.
\newblock \href {https://doi.org/10.1103/PhysRev.78.266}
  {\path{doi:10.1103/PhysRev.78.266}}.
\newline\urlprefix\url{https://link.aps.org/doi/10.1103/PhysRev.78.266}

\bibitem{Reck1969}
R.~A. Reck, D.~L. Fry,
  \href{https://link.aps.org/doi/10.1103/PhysRev.184.492}{Orbital and spin
  magnetization in {Fe-Co, Fe-Ni, and Ni-Co}}, Phys. Rev. 184 (1969) 492--495.
\newblock \href {https://doi.org/10.1103/PhysRev.184.492}
  {\path{doi:10.1103/PhysRev.184.492}}.
\newline\urlprefix\url{https://link.aps.org/doi/10.1103/PhysRev.184.492}

\bibitem{ScottSep1955}
G.~G. Scott,
  \href{https://link.aps.org/doi/10.1103/PhysRev.99.1824}{Gyromagnetic ratio of
  nickel at low magnetic intensities}, Phys. Rev. 99 (1955) 1824--1825.
\newblock \href {https://doi.org/10.1103/PhysRev.99.1824}
  {\path{doi:10.1103/PhysRev.99.1824}}.
\newline\urlprefix\url{https://link.aps.org/doi/10.1103/PhysRev.99.1824}

\bibitem{Standley1955}
K.~J. Standley, K.~H. Reich,
  \href{http://stacks.iop.org/0370-1301/68/i=10/a=303}{Ferromagnetic resonance
  in nickel and in some of its alloys}, Proceedings of the Physical Society.
  Section B 68~(10) (1955) 713.
\newline\urlprefix\url{http://stacks.iop.org/0370-1301/68/i=10/a=303}

\bibitem{borovik1988}
A.~Borovik-Romanov, S.~Sinha,
  \href{https://books.google.com/books?id=Qj9BAQAAIAAJ}{Spin Waves and Magnetic
  Excitations}, no.~2 in Modern problems in condensed matter sciences,
  North-Holland, 1988.
\newline\urlprefix\url{https://books.google.com/books?id=Qj9BAQAAIAAJ}

\bibitem{Gadsden1978}
C.~J. Gadsden, M.~Heath,
  \href{http://stacks.iop.org/0305-4608/8/i=3/a=021}{Ferromagnetic resonance of
  nickel vanadium alloys}, Journal of Physics F: Metal Physics 8~(3) (1978)
  521.
\newline\urlprefix\url{http://stacks.iop.org/0305-4608/8/i=3/a=021}

\bibitem{Shanina1998}
B.~D. Shanina, V.~G. Gavriljuk, A.~A. Konchits, S.~P. Kolesnik,
  \href{http://stacks.iop.org/0953-8984/10/i=8/a=015}{The influence of
  substitutional atoms upon the electron structure of the iron-based transition
  metal alloys}, Journal of Physics: Condensed Matter 10~(8) (1998) 1825.
\newline\urlprefix\url{http://stacks.iop.org/0953-8984/10/i=8/a=015}

\bibitem{Dewar1977}
G.~Dewar, B.~Heinrich, J.~F. Cochran,
  \href{https://doi.org/10.1139/p77-112}{Ferromagnetic antiresonance
  transmission of 24~{GH}z radiation through nickel (20 to 364 °c)}, Canadian
  Journal of Physics 55~(9) (1977) 821--833.
\newblock \href {https://doi.org/10.1139/p77-112} {\path{doi:10.1139/p77-112}}.
\newline\urlprefix\url{https://doi.org/10.1139/p77-112}

\bibitem{Pust1981}
L.~Půst, Z.~Frait,
  \href{http://www.sciencedirect.com/science/article/pii/037596018190685X}{Precise
  g-factor determination of {Fe-3wt\%Si} single crystals in the temperature
  range 3.5 - 300 {K} by electron {FMR} and {FMAR} measurements}, Physics
  Letters A 86~(1) (1981) 48 -- 50.
\newblock \href
  {https://doi.org/http://dx.doi.org/10.1016/0375-9601(81)90685-X}
  {\path{doi:http://dx.doi.org/10.1016/0375-9601(81)90685-X}}.
\newline\urlprefix\url{http://www.sciencedirect.com/science/article/pii/037596018190685X}

\bibitem{Haraldson1981}
S.~Haraldson, L.~Pettersson,
  \href{http://www.sciencedirect.com/science/article/pii/0022369781901219}{Ferromagnetic
  resonance in nickel around the curie temperature}, Journal of Physics and
  Chemistry of Solids 42~(8) (1981) 681 -- 686.
\newblock \href
  {https://doi.org/http://dx.doi.org/10.1016/0022-3697(81)90121-9}
  {\path{doi:http://dx.doi.org/10.1016/0022-3697(81)90121-9}}.
\newline\urlprefix\url{http://www.sciencedirect.com/science/article/pii/0022369781901219}

\bibitem{Bastian1976_2}
D.~Bastian, E.~Biller,
  \href{http://dx.doi.org/10.1002/pssa.2210350207}{Anisotropy constants and
  g-factors of {NiFe} alloys derived from ferromagnetic resonance}, physica
  status solidi (a) 35~(2) (1976) 465--470.
\newblock \href {https://doi.org/10.1002/pssa.2210350207}
  {\path{doi:10.1002/pssa.2210350207}}.
\newline\urlprefix\url{http://dx.doi.org/10.1002/pssa.2210350207}

\bibitem{Rodbell1964}
D.~S. Rodbell,
  \href{https://link.aps.org/doi/10.1103/PhysRevLett.13.471}{Ferromagnetic
  resonance absorption linewidth of nickel metal. {E}vidence for
  {Landau-Lifshitz} damping}, Phys. Rev. Lett. 13 (1964) 471--474.
\newblock \href {https://doi.org/10.1103/PhysRevLett.13.471}
  {\path{doi:10.1103/PhysRevLett.13.471}}.
\newline\urlprefix\url{https://link.aps.org/doi/10.1103/PhysRevLett.13.471}

\bibitem{Rodbell1959}
D.~S. Rodbell, \href{http://dx.doi.org/10.1063/1.2185880}{Ferromagnetic
  resonance of iron whisker crystals}, Journal of Applied Physics 30~(4) (1959)
  S187--S188.
\newblock \href {https://doi.org/10.1063/1.2185880}
  {\path{doi:10.1063/1.2185880}}.
\newline\urlprefix\url{http://dx.doi.org/10.1063/1.2185880}

\bibitem{Bagguley1954}
D.~M.~S. Bagguley, N.~J. Harrick,
  \href{http://stacks.iop.org/0370-1298/67/i=7/a=115}{The temperature
  dependence of ferromagnetic resonance in colloidal nickel}, Proceedings of
  the Physical Society. Section A 67~(7) (1954) 648.
\newline\urlprefix\url{http://stacks.iop.org/0370-1298/67/i=7/a=115}

\bibitem{Bloembergen1950}
N.~Bloembergen, \href{https://link.aps.org/doi/10.1103/PhysRev.78.572}{On the
  ferromagnetic resonance in nickel and supermalloy}, Phys. Rev. 78 (1950)
  572--580.
\newblock \href {https://doi.org/10.1103/PhysRev.78.572}
  {\path{doi:10.1103/PhysRev.78.572}}.
\newline\urlprefix\url{https://link.aps.org/doi/10.1103/PhysRev.78.572}

\bibitem{ScottAug1955}
G.~G. Scott,
  \href{https://link.aps.org/doi/10.1103/PhysRev.99.1241}{Gyromagnetic ratio of
  iron at low magnetic intensities}, Phys. Rev. 99 (1955) 1241--1244.
\newblock \href {https://doi.org/10.1103/PhysRev.99.1241}
  {\path{doi:10.1103/PhysRev.99.1241}}.
\newline\urlprefix\url{https://link.aps.org/doi/10.1103/PhysRev.99.1241}

\bibitem{ScottAug1956}
G.~G. Scott,
  \href{https://link.aps.org/doi/10.1103/PhysRev.103.561}{Gyromagnetic ratios
  of the iron-nickel alloys}, Phys. Rev. 103 (1956) 561--563.
\newblock \href {https://doi.org/10.1103/PhysRev.103.561}
  {\path{doi:10.1103/PhysRev.103.561}}.
\newline\urlprefix\url{https://link.aps.org/doi/10.1103/PhysRev.103.561}

\bibitem{Frait1971}
{FRAIT, Z.}, {GEMPERLE, R.},
  \href{https://doi.org/10.1051/jphyscol:19711182}{The g-factor and surface
  magnetization of pure iron along [100] and [111] directions}, J. Phys.
  Colloques 32 (1971).
\newblock \href {https://doi.org/10.1051/jphyscol:19711182}
  {\path{doi:10.1051/jphyscol:19711182}}.
\newline\urlprefix\url{https://doi.org/10.1051/jphyscol:19711182}

\bibitem{Kittel1948}
C.~Kittel, \href{https://link.aps.org/doi/10.1103/PhysRev.73.155}{On the theory
  of ferromagnetic resonance absorption}, Phys. Rev. 73 (1948) 155--161.
\newblock \href {https://doi.org/10.1103/PhysRev.73.155}
  {\path{doi:10.1103/PhysRev.73.155}}.
\newline\urlprefix\url{https://link.aps.org/doi/10.1103/PhysRev.73.155}

\bibitem{Frait1977}
Z.~Frait, \href{https://doi.org/10.1007/BF01587010}{The g-factor in pure
  polycrystalline iron}, Czechoslovak Journal of Physics B 27~(2) (1977)
  185--189.
\newblock \href {https://doi.org/10.1007/BF01587010}
  {\path{doi:10.1007/BF01587010}}.
\newline\urlprefix\url{https://doi.org/10.1007/BF01587010}

\bibitem{Frait1965}
Z.~Frait, H.~MacFaden,
  \href{https://link.aps.org/doi/10.1103/PhysRev.139.A1173}{Ferromagnetic
  resonance in metals. {F}requency dependence}, Phys. Rev. 139 (1965)
  A1173--A1181.
\newblock \href {https://doi.org/10.1103/PhysRev.139.A1173}
  {\path{doi:10.1103/PhysRev.139.A1173}}.
\newline\urlprefix\url{https://link.aps.org/doi/10.1103/PhysRev.139.A1173}

\bibitem{Herring1966}
C.~Herring, R.~M. Bozorth, A.~E. Clark, T.~R. McGuire,
  \href{https://doi.org/10.1063/1.1708462}{High‐field susceptibilities of
  iron and nickel}, Journal of Applied Physics 37~(3) (1966) 1340--1341.
\newblock \href {http://arxiv.org/abs/https://doi.org/10.1063/1.1708462}
  {\path{arXiv:https://doi.org/10.1063/1.1708462}}, \href
  {https://doi.org/10.1063/1.1708462} {\path{doi:10.1063/1.1708462}}.
\newline\urlprefix\url{https://doi.org/10.1063/1.1708462}

\bibitem{Stoelinga1966}
J.~Stoelinga, R.~Gersdorf, Field dependence of the magnetization in high fields
  for bcc fe-co and fe-ni alloys, Physics Letters 19~(8) (1966) 640--641.

\bibitem{Yasui1971}
M.~Yasui, M.~Shimizu, \href{https://doi.org/10.1143/JPSJ.31.378}{Calculations
  of orbital paramagnetic susceptibility for vanadium, chromium and iron},
  Journal of the Physical Society of Japan 31~(2) (1971) 378--381.
\newblock \href {http://arxiv.org/abs/https://doi.org/10.1143/JPSJ.31.378}
  {\path{arXiv:https://doi.org/10.1143/JPSJ.31.378}}, \href
  {https://doi.org/10.1143/JPSJ.31.378} {\path{doi:10.1143/JPSJ.31.378}}.
\newline\urlprefix\url{https://doi.org/10.1143/JPSJ.31.378}

\bibitem{Foner1966}
A.~J. Freeman, N.~A. Blum, S.~Foner, R.~B. Frankel, E.~J. McNiff,
  \href{https://doi.org/10.1063/1.1708461}{Ferromagnetic metals in high
  magnetic fields}, Journal of Applied Physics 37~(3) (1966) 1338--1339.
\newblock \href {http://arxiv.org/abs/https://doi.org/10.1063/1.1708461}
  {\path{arXiv:https://doi.org/10.1063/1.1708461}}, \href
  {https://doi.org/10.1063/1.1708461} {\path{doi:10.1063/1.1708461}}.
\newline\urlprefix\url{https://doi.org/10.1063/1.1708461}

\end{thebibliography}
\end{document}